\documentclass[
  journal=pasa,
  manuscript=research-paper, 
  year=2020,
  volume=37,
]{cup-journal}

\usepackage{microtype,siunitx,booktabs}
\usepackage[utf8]{inputenc}
\usepackage{caption}
\usepackage{subcaption}
\usepackage{gensymb}

\title{Localisation of gamma-ray bursts from the combined SpIRIT+HERMES-TP/SP nano-satellite constellation}

\author{M. Thomas}
\affiliation{School of Physics, The University of Melbourne, VIC 3010, Australia}
\email[M. Thomas]{thomasm3@student.unimelb.edu.au}

\author{M. Trenti}
\affiliation{School of Physics, The University of Melbourne, VIC 3010, Australia}
\alsoaffiliation{Australian Research Council Centre of Excellence for All-Sky Astrophysics in 3-Dimensions, Australia}

\author{A. Sanna}
\affiliation{Dipartimento di Fisica, Universit\`{a} degli Studi di Cagliari, SP Monserrato-Sestu km 0.7, I-09042 Monserrato, Italy}

\author{R. Campana}
\affiliation{INAF - Osservatorio di Astrofisica e Scienza dello Spazio, Via Gobetti 93/3, 40129 - Bologna, Italy}

\author{G. Ghirlanda}
\affiliation{INAF - Osservatorio Astronomico di Brera, Via E. Bianchi 46, 23807 Merate (LC), Italy}
\alsoaffiliation{INFN – Sezione di Milano-Bicocca, Piazza della Scienza 3, 20126 Milano, Italy}

\author{J. {\v R}{\'i}pa}
\affiliation{Masaryk University, Faculty of Science, Department of Theoretical Physics and Astrophysics, Kotl\'{a}\v{r}sk\'{a} 2, Brno, Czech Republic, 611 37}

\author{L. Burderi}
\affiliation{Dipartimento di Fisica, Universit\`{a} degli Studi di Cagliari, SP Monserrato-Sestu km 0.7, I-09042 Monserrato, Italy}

\author{F. Fiore}
\affiliation{INAF - Osservatorio di Astrofisica e Scienza dello Spazio, Via Gobetti 93/3, 40129 - Bologna, Italy}

\author{Y. Evangelista}
\affiliation{INAF - Istituto di Astrofisica e Planetologia Spaziali, Via del Fosso del Cavaliere, 100, 00133 Roma RM, Italy}
\alsoaffiliation{INFN - Sezione Tor Vergata, Via della Ricerca Scientifica, I-00133 Rome, Italy}

\author{L. Amati}
\affiliation{INAF - Osservatorio di Astrofisica e Scienza dello Spazio, Via Gobetti 93/3, 40129 - Bologna, Italy}

\author{S. Barraclough}
\affiliation{School of Physics, The University of Melbourne, VIC 3010, Australia}

\author{K. Auchettl}
\affiliation{OzGrav, School of Physics, The University of Melbourne, Parkville, Victoria 3010, Australia.}
\alsoaffiliation{Australian Research Council Centre of Excellence for All-Sky Astrophysics in 3-Dimensions, Australia}
\alsoaffiliation{Department of Astronomy and Astrophysics, University of California, Santa Cruz, CA 95064, USA}

\author{M. O. del Castillo}
\affiliation{School of Physics, The University of Melbourne, VIC 3010, Australia}

\author{A. Chapman}
\affiliation{Department of Mechanical Engineering, School of Engineering, The University of Melbourne, Australia}

\author{M. Citossi}
\affiliation{INAF - Osservatorio Astronomico di Trieste, Via GB Tiepolo, 11 I-34143 Trieste, Italy}

\author{A. Colagrossi}
\affiliation{Department of Aerospace Science and Technology, Politecnico di Milano, Via La Masa 34, 20156 Milano, Italy}

\author{G. Dilillo}
\affiliation{INAF - Istituto di Astrofisica e Planetologia Spaziali, Via del Fosso del Cavaliere, 100, 00133 Roma RM, Italy}

\author{N. Deiosso}
\affiliation{Dipartimento di Fisica, Universit\`{a} degli Studi di Cagliari, SP Monserrato-Sestu km 0.7, I-09042 Monserrato, Italy}

\author{E. Demenev}
\affiliation{FBK - Sensors and Devices, Micro Nano Facility, via Sommarive, 18 38123 Povo, Trento, Italy}

\author{F. Longo}
\affiliation{Department of Physics, University of Trieste, via Valerio 2, Trieste, Italy}
\alsoaffiliation{INFN, sezione di Trieste, via Valerio 2, Trieste, Italy}
\alsoaffiliation{Institute for Fundamental Physics of the Universe (IFPU), via Beirut 2, Trieste, Italy}

\author{A. Marino}
\affiliation{Institute of Space Sciences (ICE, CSIC), Campus UAB, Carrer de Can Magrans s/n, E-08193 Barcelona, Spain}
\alsoaffiliation{Institut d'Estudis Espacials de Catalunya (IEEC), E-08034 Barcelona, Spain}
\alsoaffiliation{INAF/IASF Palermo, via Ugo La Malfa 153, I-90146 - Palermo, Italy}

\author{J. McRobbie}
\affiliation{School of Physics, The University of Melbourne, VIC 3010, Australia}

\author{R. Mearns}
\affiliation{School of Physics, The University of Melbourne, VIC 3010, Australia}

\author{A. Melandri}
\affiliation{INAF - Osservatorio Astronomico di Brera, Via E. Bianchi 46, 23807 Merate (LC), Italy}

\author{A. Riggio}
\affiliation{Dipartimento di Fisica, Universit\`{a} degli Studi di Cagliari, SP Monserrato-Sestu km 0.7, I-09042 Monserrato, Italy}
\alsoaffiliation{INAF/IASF Palermo, via Ugo La Malfa 153, I-90146 - Palermo, Italy}

\author{T. Di Salvo}
\affiliation{Universit\`{a} degli Studi di Palermo, Dipartimento di Fisica e Chimica, via Archirafi 36 - 90123 Palermo, Italy}

\author{S. Puccetti}
\affiliation{ASI - Agenzia Spaziale Italiana, Via del Politecnico s.n.c., 00133 Roma, Italy}

\author{M. Topinka}
\affiliation{INAF – Istituto di Astrofisica Spaziale e Fisica Cosmica, Via A. Corti 12, I-20133 Milano, Italy}

\doi{10.1017/pasa.2020.32}

\received {dd Mmm YYYY}
\revised  {dd Mmm YYYY}
\accepted {dd Mmm YYYY}
\published{dd Mmm YYYY}


\begin{document}

\begin{abstract}
    Multi-messenger observations of the transient sky to detect cosmic explosions and counterparts of gravitational wave mergers critically rely on orbiting wide-FoV telescopes to cover the wide range of wavelengths where atmospheric absorption and emission limit the use of ground facilities. Thanks to continuing technological improvements, miniaturised space instruments operating as distributed-aperture constellations are offering new capabilities for the study of high energy transients to complement ageing existing satellites. In this paper we characterise the performance of the upcoming joint SpIRIT and HERMES-TP/SP constellation for the localisation of high-energy transients through triangulation of signal arrival times. SpIRIT is an Australian technology and science demonstrator satellite designed to operate in a low-Earth Sun-synchronous Polar orbit that will augment the science operations for the equatorial HERMES-TP/SP constellation. In this work we simulate the improvement to the localisation capabilities of the HERMES - TP/SP constellation when SpIRIT is included in an orbital plane nearly perpendicular (inclination = 97.6\degree) to the HERMES - TP/SP orbits. For the fraction of GRBs detected by three of the HERMES satellites plus SpIRIT, we find that the combined constellation is capable of localising $60\%$ of long GRBs to within $\sim 30$\,deg$^2$ on the sky, and $60\%$ of short GRBs within $\sim 1850$\,deg$^2$ ($1\sigma$ confidence regions), though it is beyond the scope of this work to characterise or rule out systematic uncertainty of the same order of magnitude. Based purely on statistical GRB localisation capabilities (i.e., excluding systematic uncertainties and sky coverage), these figures for long GRBs are comparable to those reported by the Fermi Gamma Burst Monitor instrument. These localisation statistics represents a reduction of the uncertainty for the burst localisation region for both long and short GRBs by a factor of $\sim 5$ compared to the HERMES-TP/SP alone. Further improvements by an additional factor of $2$ (or $4$) can be achieved by launching an additional 4 (or 6) SpIRIT-like satellites into a Polar orbit respectively, which would both increase the fraction of sky covered by multiple satellite elements, and also enable localisation of $\geq 60\%$ of long GRBs to within a radius of $\sim 1.5\degree$ (statistical uncertainty) on the sky, clearly demonstrating the value of a distributed all-sky high energy transient monitor composed of nano-satellites.
    
\end{abstract}

\section{Introduction}
\label{sec:Introduction}
    
    The ability to localise gamma ray bursts (GRBs) from their prompt X-ray/gamma-ray emission is a critical step in the identification and follow-up observation of these cosmic explosions. GRB afterglows are so faint that until the advent of large-aperture, wide-field, deep optical/infrared imaging telescopes (e.g., the Vera Rubin Observatory - albeit with a limited FoV; see \citealt{LSST_2017}), the location of the burst must be determined purely from its prompt high energy photon emission. 
    
    Accurate GRB localisations and subsequent afterglow follow-up observations are desirable across many areas of astronomy. Observations of long GRB afterglows offer many opportunities to study the high redshift universe, and can be used in the study of the cosmic star formation rate (e.g., \citealt{Robertson_2012, Trenti_2013, Petrosian_2015}, \citealt{Chary_2016, Lloyd-Ronning_2019}), investigating the luminosity function of high redshift galaxies (e.g., \citealt{Trenti_2012, Tanvir_2012}; \citealt{Salvaterra_2012}; \citealt{McGuire_2016}), characterising the interstellar medium in distant galaxies (e.g., \citealt{Klose_2004, Berger_2005}; \citealt{ Vreeswijk_2007, Fox_2008}; \citealt{Prochaska_2008,Thone_2013, Wiseman_2017}) and measuring the neutral hydrogen fraction along the line of sight (e.g., \citealt{Miralda_Escude_1998, Mesinger_2008, McQuinn_2008, Hartoog_2015}; \citealt{Melandri_2015, Lidz_2021}). 
    
    Furthermore, observations of prompt and afterglow emission from both long and short GRBs provide unique insights into the physics of energy transport and dissipation within relativistic jets (see e.g., \citealt{Kumar_2015, Miceli_2022} for review articles). For example, the very recent opening of the so-called VHE (Very High Energy) window, with the detection of TeV emission from GRB afterglows \citep{MAGIC_2019, Abdalla_2019, HESS_2021}, provides us with a new tool to understand the physics of particle acceleration in weakly magnetised ultra-relativistic shocks, and may in future also enable a better understanding of highly magnetised mildly relativistic shocks with VHE observations of GRB prompt emission.
    
    More recently, GRB localisations have become useful tools in the field of gravitational wave (GW) science due to the association between short GRBs and compact object mergers \citep{Eichler_1989, Perna_2002, Belczynski_2006}. The nearly simultaneous detection of a short GRB within the same localisation region as a GW event provides unique insight into the true nature of these high energy explosions, as illustrated by the simultaneous observations of the binary neutron star (BNS) coalescence event GW170817, and the short gamma-ray burst GRB170817A \citep{LIGO_2017} which solidified the link between short GRBs and binary neutron star mergers. In the future, accurate localisations of short GRBs from their high energy photon emission may also assist in more precise localisation of GW transient events in the case where the uncertainty region from the GW signal is large.
    
    A well-established method to localise high energy EM transients is triangulation, whereby the source position in the sky is determined by measuring the arrival time difference of light at each detector (see, e.g., \citealt{Hurley_2013}). This method has been used to localise GRBs since they were first discovered by the Vela satellites in 1969 \citep{Klebesadel_1973}, and has been used for the last several decades since 1976 by the Inter-Planetary Network (IPN) (NASA/GSFC Interplanetary Gamma-Ray Burst Timing Network\footnote{\url{https://heasarc.gsfc.nasa.gov/docs/heasarc/missions/ipn.html}}), an evolving group of independent, separately executed satellite missions all carrying high energy photon detectors, the data from which can be combined in order to accurately localise GRBs and other high energy transients \citep{Hurley_1999, Hurley_2013, Hurley_2017}.
    
    Due to the cost-effectiveness of CubeSat technology in achieving specific science goals and the potential benefit to the field of GRB science, many nano-satellite missions and mission concepts are under development to take advantage of this emergent new technology. Such missions include the GRID mission \citep{Wen_2019, Wang_2021b}, CAMELOT \citep{Werner_2018}, GECAM \citep{Zhang_2019}, BurstCube \citep{Racusin_2017}, EIRSAT-1 \citep{Murphy_2021}, and MoonBEAM \citep{Hui_2021}. Similarly, small hosted payload instruments on a constellation of larger satellites have been proposed \citep{Greiner_2022}. 

    When combined together, all these upcoming instruments for high energy astrophysics have the potential to significantly contribute to and improve the existing network of GRB satellites within the next decade. In fact, they could establish an all-sky X/gamma-ray monitor capable of high precision localisations and low-latency communications with the ground (given many elements would be in low Earth orbit), which would dramatically increase the number of GRBs identified in near-real time and thus accessible for rapid afterglow follow-up observations.
    
    The HERMES - TP/SP (High Energy Rapid Modular Ensemble of Satellites - Technologic/Scientific Pathfinder) is a constellation of six 3U nano-satellites (the size of small standardized-form-factor satellites is measured in `Units' U, where 1U $\approx 10 \times 10 \times 10$\,cm$^3$), hosting innovative wide field-of-view (FOV) X-ray detectors for the monitoring and localisation of GRBs and high energy transients \citep{Fiore_2020}. Each satellite will be launched into an equatorial low Earth orbit (LEO), and the constellation is designed to not only triangulate GRBs and high-energy transients on the sky \citep{Sanna_2020} but also to probe the quantum structure of spacetime thanks to sub $\mu s$ timing resolution of the HERMES instrument \citep{Burderi_2020}. The HERMES - TP/SP is designed to demonstrate accurate GRB localisation mainly in the right ascension coordinate - the equatorial orbit of the pathfinder satellites imposes inherent geometrical limitations on triangulation which means that GRBs will be only loosely constrained in the declination coordinate \citep{Sanna_2020}. This limitation is expected to be addressed in a future HERMES Full Constellation (FC), which is envisioned to consist of tens of satellites into several different orbital planes to achieve accuracy of localising bright long GRBs better than $15'$ \citep{Fiore_2020}.
    
    As a first step to demonstrate the potential of the HERMES FC, the SpIRIT (Space Industry Responsive Intelligent Thermal) nano-satellite is an Australia-Italy nano-satellite mission planned for launch in 2023 which will operate as part of the HERMES - TP/SP, forming a combined constellation of seven satellites designed to detect and localise high energy transient events. SpIRIT will be the only satellite among the HERMES Pathfinder to be launched into a Polar orbit, which will enable improved localisation in the declination coordinate due to SpIRIT's large baseline perpendicular to the equatorial plane. In this capacity, SpIRIT will act as a proof-of-concept for the HERMES FC to demonstrate the GRB localisation capabilities with satellites in different orbital planes.
    
    In this paper we simulate the localisation capabilities of the combined SpIRIT + HERMES - TP/SP constellation in order to understand the improvement gained by having a single satellite element in Polar orbit. As an extension, we further investigate the localisation capabilities of a combined constellation with as many as six additional satellites launched into Polar orbit.
    
    This paper is organised as follows: Section \ref{sec:HERMES - TP/SP} provides a brief background to the HERMES Technologic/Scientific Pathfinder mission and gives details regarding the HERMES X-ray instrument. Section \ref{sec:SpIRIT Nano-Satellite} introduces the SpIRIT nano-satellite mission. Section \ref{sec:Simulating GRB Triangulation} describes our simulation framework and modelling assumptions, as well as the mathematical techniques used for the triangulation and localisation of GRBs. In Section \ref{sec:Results and Discussion} we compare the GRB localisation capabilities and the sky coverage statistics of the HERMES - TP/SP with and without the SpIRIT satellite, and  with a larger number of polar nano-satellites. The main conclusions from this work are presented in Section \ref{sec:Conclusion}.
    
\section{HERMES-Technologic and Scientific Pathfinder}
\label{sec:HERMES - TP/SP}

    The HERMES - TP/SP is a constellation of six 3U nano-satellites being developed with funding from the Italian Space Agency and the Italian Ministry for education, University and Research (PI: Burderi) and from the European Union’s Horizon 2020 Research and Innovation Program (PI Fiore; see \citet{Fiore_2020} for further details). The HERMES - TP/SP is an in-orbit demonstration of the capabilities of distributed aperture astronomy in the field of GRB science. It is designed as an intrinsically modular experiment which can later be naturally expanded to provide a global, sensitive all-sky monitor for high-energy transients (the HERMES - FC; see e.g., \citealt{Fuschino_2019} for discussion).
    
    The HERMES - TP/SP mission is in advanced stages of development. The critical design review was successfully passed in late 2020 \citep{Fiore_2020}, and integration and testing of flight units has been taking place since mid-2021. Integration and testing is continuing throughout 2022, and the constellation is expected to be launched to a nearly equatorial (inclination $\leq 20\degree$) low Earth orbit \citep{Fiore_2021b}.
    
    Each satellite hosts a miniaturised X-ray detector - the HERMES instrument - based on the so-called `siswich' concept of silicon drift detectors (SDD's) coupled with scintillator crystals. The instrument has a burst sensitivity of 2 photons/cm$^2$/s for $E \leq 20$keV and 1 photon/cm$^2$/s for $50 \leq E \leq 300$keV (which is comparable to the Fermi GBM burst sensitivity of $< 0.5$ photons/cm$^2$/s between 50-300 keV (NASA/GSFC Fermi Gamma-ray Space Telescope GBM Specifications \& Performance\footnote{\url{https://fermi.gsfc.nasa.gov/science/instruments/table1-2.html}})), and low background for its size \citep{Evangelista_2020a, Campana_2020}. Its two operative modes  -- direct X-ray absorption on the SDD or scintillation light readout from the GAGG crystals -- cover a wide energy range (3-60 keV and 20-2000 keV, respectively; \citealt{Fuschino_2020}). The instrument has excellent temporal resolution ($<400$ns), and has been designed for detection of cosmic high energy transients (such as GRBs) and for the determination of their positions through triangulation via the distributed detector architecture. 
    
    The response function of the HERMES instrument depends on both the energy of the incoming photon and its incident angle on the detector relative to the line of sight (LOS), and we refer to \citet{Campana_2020} for a plot of the instrument response as a function of photon energy and incident angle. The response of the HERMES instrument as a function of incident angle for energies between $50-300$\,keV can be approximated by a cosine profile with respect to the LOS, meaning that the full width at half maximum (FWHM) of the instrument is 120deg, corresponding to an effective FOV of 3.2sr FWHM or, equivalently, a circular FOV with a radius $60\degree$ \citep{Colagrossi_2020, Evangelista_2020a}. In this work we adopt the use of a cosine profile out to a radius of 80 degrees for the instrument response (IR):
    \subsubsection*{}
    
        \begin{equation}\label{eq:HERMES instrument response}
            IR(\theta) =
            \begin{cases}
                \cos\theta &,\theta<80\degree \\
                0 &,\theta>80\degree \\
            \end{cases} \\
        \end{equation} \\
    
    The 80 degree cutoff is motivated as approximately representative of the physical obstructions to the detector field of view, given how the instrument is mounted on the SpIRIT satellite. We note that we have assumed that the aluminum chassis of the satellite and the tungsten shielding of the instrument block all radiation coming from incident angles $> 80\degree$ - in reality this may not be the case, and a more detailed procedure -- or ultimately in-orbit verification against GRBs with known localisation -- would be needed to determine which satellites observe the GRB within their FOV and which observe it off-axis. In the context of this work, the approximation given in Equation \ref{eq:HERMES instrument response} is sufficient for quantifying the improvement in GRB localisations gained by augmenting the HERMES - TP/SP with satellites in polar orbit.

\section{SpIRIT Nano-Satellite} \label{sec:SpIRIT Nano-Satellite}

    The SpIRIT satellite is an Australia-Italy mission supported in Australia by the Australian Space Agency International Space Investment - Expand Capability scheme. SpIRIT is a 6U CubeSat with $\sim 11.5~\mathrm{kg}$ mass and linear dimensions of approximately $30\times 20 \times 10~\mathrm{cm}$ when stowed in the launch dispenser. SpIRIT will be launched in a Polar Sun-synchronous orbit with $\sim 550~\mathrm{km}$ altitude and approximately 1:30pm LTDN (Local Time at the Descending Node), and has a target main mission lifetime of 24 months including commissioning. 
    
    \begin{figure}[t!]
        \centering
        \includegraphics[width = 0.8\textwidth]{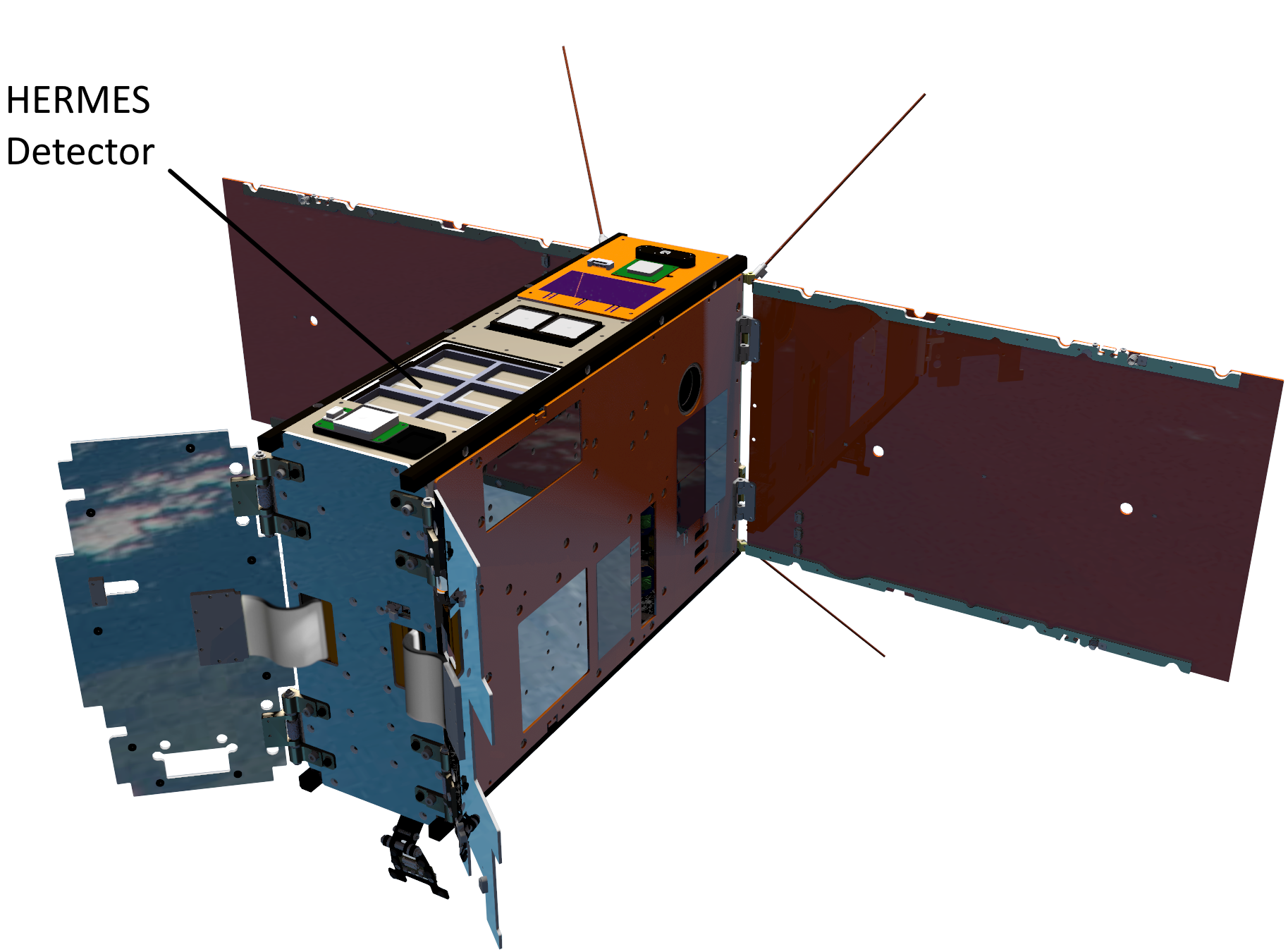}
        \caption{Diagram of the SpIRIT nano-satellite, demonstrating the orientation of the solar panels (facing away from the view on the right) relative to the HERMES instrument (located on the top face near the left). Also visible on the left side are the deployable thermal radiators. 
        }
        \label{fig:spirit reference image}
    \end{figure}
    
    SpIRIT is being developed by an Australian consortium led by the University of Melbourne (PI: Michele Trenti), which is responsible for the project management and mission design, as well as for the development of four subsystems: 
    
    \begin{enumerate}
    
        \item The Payload Management System (PMS), a dedicated computer that manages operations of all payloads and related data processing. This system is based on a SmartFusion2 System on Chip (SoC) Field Programmable Gate Array (FPGA) which includes a Cortex-M3 microprocessor and FPGA logic. A Graphic Processing Unit is also part of PMS for additional on-board data processing capabilities (including a progressive compression algorithm for improved performance in imaging data downlink). PMS also includes a Power Management Module with multiple supercapacitors to guarantee up to 60 seconds of uninterrupted power supply to the HERMES instrument and allow it to transition to a safe mode in case of temporary power failure at the platform level.
        
        \item The Thermal Management Integrated System (TheMIS), a subsystem for precision thermal management and active cooling of payloads that includes (a) a Stirling cycle cryocooler capable of achieving $T=80~\mathrm{K}$ cold-tip temperature, (b) custom-developed control electronics designed for space operations, (c) multiple temperature monitoring sensors, (d) and deployable thermal radiators. 
        
        \item Mercury, a low-latency communication subsystem that combines Iridium and Globalstar user terminals to provide capability of sending and receiving short burst data packets (e.g. target of opportunity commands or burst alerts).
        
        \item LORIS, a set of visible and near-infrared inspection cameras that will monitor and characterise key elements of the spacecraft, including one mounted on a deployable arm, seen on the bottom of the satellite in Figure \ref{fig:spirit reference image}. 
    \end{enumerate}
 
    The main scientific payload of SpIRIT is a HERMES instrument unit, identical to the ones that will fly on the HERMES - TP/SP satellites. The HERMES instrument on SpIRIT will accumulate a higher total ionizing dose compared to a satellite in an equatorial orbit due to its repeated passage over the South Atlantic Anomaly and the Poles. To reduce the resultant buildup of leakage current and mitigate the performance degradation of the instrument \citep{Dilillo_2020}, SpIRIT will feature active thermal management conducted by the TheMIS payload, with a target operating temperature of $T=245\pm1~\mathrm{K}$. The lower the temperature the greater the benefit, and TheMIS would, in principle, be capable of cooling a well-insulated instrument to cryogenic temperatures ($T\sim 80~\mathrm{K}$), but the HERMES instrument electronics in the current design are only rated to $T\gtrsim 240~\mathrm{K}$. A CAD illustration of the satellite is shown in Figure~\ref{fig:spirit reference image}.

    All of these elements make SpIRIT a mission designed to demonstrate new technologies in orbit and the scientific operation and performance of the HERMES instrument in high inclination orbits.

\section{Simulating GRB Triangulation}
\label{sec:Simulating GRB Triangulation}

In this Section we introduce our analysis framework to simulate and localise GRBs using the HERMES (+ SpIRIT) nanosatellite constellation. 

    \subsection{Satellite Orbits}
    \label{sec:Satellite Orbits}
       
        Using as a starting point of our analysis the orbital configuration and pointing strategy presented in \citet{Colagrossi_2020}, we simulate each of the six satellites in the HERMES - TP/SP as being uniformly spaced around the same circular 550km equatorial orbit ($\text{inclination} = 0$, $\text{eccentricity} = 0$). 
        
        For simplicity we presume that the satellites maintain their equidistant spacing over time, neglecting environmental perturbations to each spacecraft's orbit. We note that this configuration is close to the optimal configuration of the HERMES - TP/SP as it maintains a large baseline between neighbouring satellites, meaning that our results will show an idealised comparison between the HERMES - TP/SP and SpIRIT + HERMES - TP/SP constellations.
        
        We model the SpIRIT satellite in a 550km Polar sun-synchronous circular orbit ($\text{inclination} = 97.6\degree$, $\text{eccentricity} = 0$) with a 1:30pm LTDN. When modelling several SpIRIT-like satellites in Polar orbit, we position them equally around a 550km Polar sun-synchronous orbit, just as the HERMES-TP/SP is uniformly spaced around an equatorial orbit. The spacing of the satellites around a Polar orbit depends on how many satellites are in this orbital plane, e.g., two satellites in Polar orbit will be on opposite sides of the globe, but four satellites will only be a quarter of the way around the globe from their nearest neighbour.

        \begin{figure}[t!]
            \begin{center}
            \includegraphics[width=\textwidth]{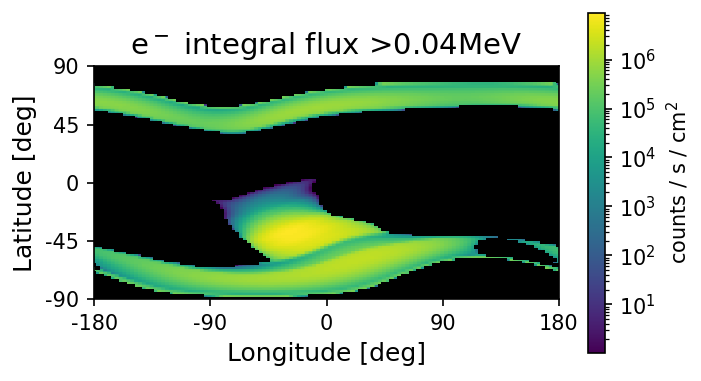}
            \includegraphics[width = \textwidth]{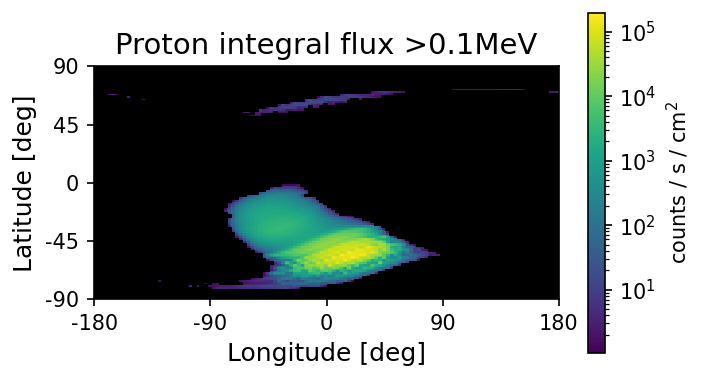}
            \caption{Particle background maps (in Earth coordinates) for a satellite in 550km low Earth orbit, obtained using the AP8MIN (protons) and AE8MAX (electrons) models included in ESA's SPENVIS system. Note that the South Atlantic Anomaly (the high particle flux region around lat $\sim -45$, lon $\sim 0$) features in both the electron and proton flux maps. The high-flux bars across the top and bottom of the e$^-$ map reflect the rings around Earth's poles. In the colour scheme, black indicates regions with zero integral flux.
            }
            \label{fig:Particle Background Maps}
            \end{center}
        \end{figure}
        
    \subsection{Particle Background Modeling}
    \label{sec:Background}
    
        The scientific operations of the HERMES and SpIRIT satellites are highly dependent on the high energy particle flux in low-Earth orbit. Space radiation can not only damage the solid-state sensors \citep{Ripa_2020}, but it also increases the level of background noise in the HERMES instrument \citep{Dilillo_2020, Dilillo_2022}, making it more difficult to detect faint GRBs and increasing the signal cross-correlation uncertainty when trying to triangulate a burst. While the influence of high energy particles is minimal for satellites in low-altitude equatorial orbit, they will have a substantial impact on SpIRIT which -- due to its Polar orbit -- will traverse multiple times a day both of the high flux regions around the poles as well as the South Atlantic Anomaly (SAA). While there is still a background flux of particles outside of the SAA and the poles (see e.g., \citealt{Ling_1975, Cumai_2019}), the flux is low enough that it does not prohibit the detection of GRBs or pose a risk to the operation of the detector.
        
        In this work we make use of particle background maps from ESA's Space Environment Information System\footnote{\url{https://www.spenvis.oma.be/credits.php}} (SPENVIS), specifically the time-averaged electron integral flux above 0.04MeV and the proton integral flux above 0.1MeV obtained using the AP8MIN and AE8MAX models included in SPENVIS (Figure \ref{fig:Particle Background Maps}). Conservatively, we assume that SpIRIT and the HERMES satellites will not operate the HERMES instrument while travelling through a region with non-zero particle background in the maps considered (non-black regions in Figure \ref{fig:Particle Background Maps}), and we take this into account in the orbital simulations presented in Section \ref{sec:Simulation Tools}. With this assumption, and taking into account the orbits defined in Section \ref{sec:Satellite Orbits}, we find that on average each satellite in the HERMES - TP/SP has approximately a $\sim 92\%$ duty cycle, while SpIRIT (and any satellite in Polar orbit) has a $\sim 65\%$ duty cycle. 
        
        As a consistency check for the duty cycle and particle background modeling, we considered in-flight data from VZLUSAT-2 \citep{Granja_2022} (mission overview available online at the VZLUSAT-2 website\footnote{ \url{https://www.vzlusat2.cz/}.}) and GRBAlpha \citep{Pal_2020}, two satellites that carry a similar detector to the HERMES instrument. Here we computed the fraction of their orbits where the instrument noise is below the threshold to detect a typical long GRB at $S/N>5$ (i.e. assuming a GRB flux greater than $1.19~\mathrm{ph/cm^2/s}$ for energy $E>30~\mathrm{keV}$) using the simulation framework outlined in \citet{Galgoczi_2021}. After mapping those results to the expected orbit of SpIRIT (taking advantage of the similarity with VZLUSAT-2, which is also in a Polar orbit, and rescaling for the sensitivity of HERMES) we estimate a 67\% duty cycle for SpIRIT, which is in very good agreement with the estimate derived using the particle background maps.    
        
        \begin{figure*}[t!]
            \begin{center}
            \includegraphics[width=0.45\textwidth]{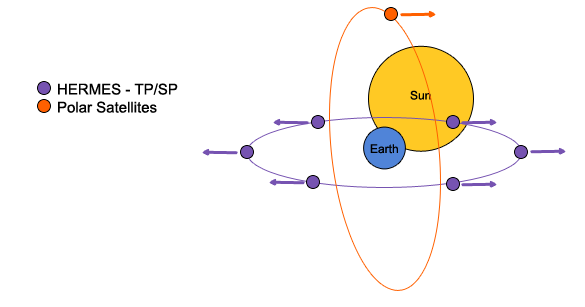}
            \includegraphics[width = 0.45\textwidth]{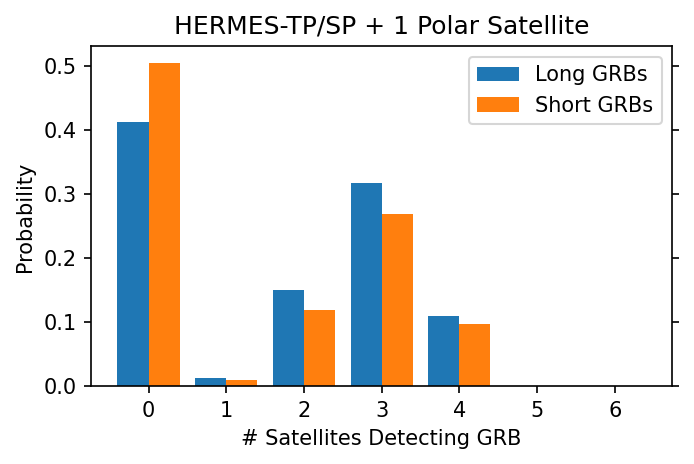}
            \\
            \includegraphics[width=0.45\textwidth]{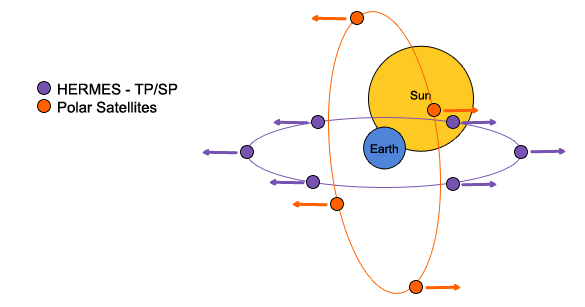}
            \includegraphics[width = 0.45\textwidth]{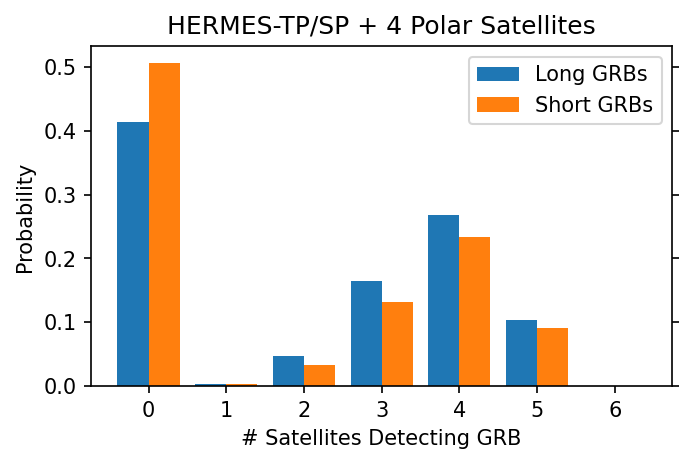}
            \\
            \includegraphics[width=0.45\textwidth]{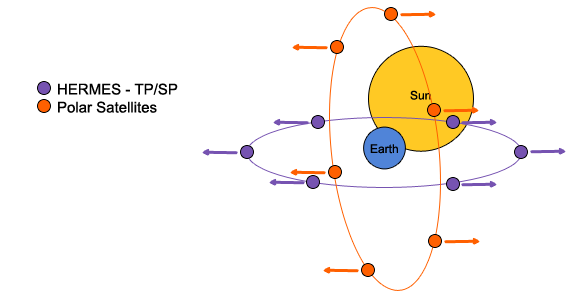}
            \includegraphics[width = 0.45\textwidth]{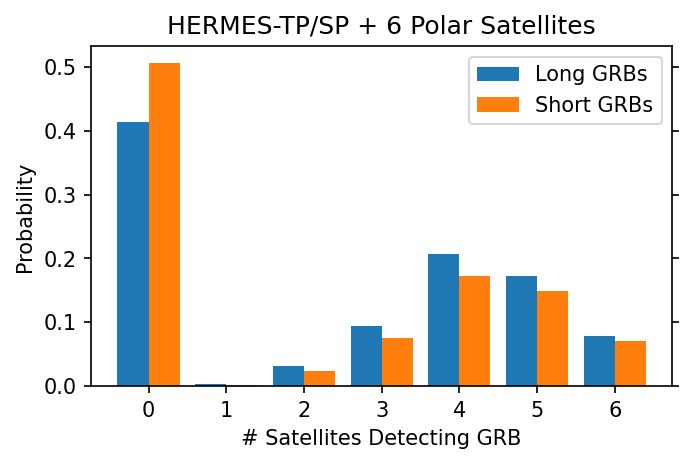}
            \caption{\textit{Left column}: 3D representation of the pointing strategy of the combined SpIRIT + HERMES - TP/SP constellation (not to scale). Note that for each constellation the satellites in a given orbital plane are uniformly spaced around their orbit. Note that for a satellite with $n$ elements in Polar orbit, the maximum number of Polar-orbiting satellites which can simultaneously detect a GRB is $n/2$ (with the exception of the case of a single Polar satellite). \textit{Right column}: Sky-coverage analysis of the corresponding constellation demonstrating the probability that a random GRB will be detected by $n$ satellites (taking into account satellite FOV, the nominal flux limit of the HERMES instrument, and satellite passage through regions of high particle flux) generated from $10^4$ simulated trials.}
            \label{fig:Constellation Pointing & Visibility}
            \end{center}
        \end{figure*}
        
    \subsection{Satellite Pointing Strategy}
    \label{sec:Pointing Strategy}
    
        The pointing direction of SpIRIT is constrained by several factors. The first constraint is the requirement to keep its solar panels illuminated by the Sun (i.e. within $\pm 20\degree$ of the Sun normal vector), which in turn restricts the direction that the on board HERMES detector - located on a face orthogonal to the solar panel -  can point at, since the solar panels are fixed (see Figure \ref{fig:spirit reference image}). Secondly, it is undesirable to point the HERMES instrument within a $90\degree$ angle of the Sun, as this would cause the detector to be directly illuminated by sunlight. The final constraint that we impose on SpIRIT's pointing in this work is the requirement to align its field of view with that of the HERMES - TP/SP to ensure the concurrent detection of GRB events.
        
        A full optimisation of the pointing strategy of a nano-satellite constellation which combines satellites in equatorial and Polar orbits with additional astrophysical constraints (i.e. preference for avoidance of the galactic plane) is beyond the scope of this work. Instead, we adopt a simplified framework (Figure \ref{fig:Constellation Pointing & Visibility}) taking advantage of the pointing strategy optimisation already carried out by the HERMES - TP/SP team \citep{Colagrossi_2020, Sanna_2020}:
        
        \begin{itemize}
            
            \item The six satellites in the HERMES - TP/SP follow the optimal pointing strategy outlined in \citet{Sanna_2020}, where the constellation is arranged into two co-pointing groups of three satellites. In this work we orient those two co-pointing groups at right angles to the sun, towards the coordinates $(\alpha, \delta) = (\alpha_{\text{sun}} + 90\degree, 0\degree)$ or $(\alpha_{\text{sun}}-90\degree,0\degree)$. While the satellites orbit around the globe they will re-orient themselves to point at whichever field of view is not obstructed by the Earth. This results in both pointing directions being constantly observed by three HERMES - TP/SP satellites at any given time.
            
            \item The SPIRIT nano-satellite will align its field of view with one of the HERMES - TP/SP triplets at any given point in time, pointing at either  $(\alpha, \delta) = (\alpha_{\text{sun}} + 90\degree, 0\degree)$ or $(\alpha_{\text{sun}}-90\degree,0\degree)$. As SpIRIT crosses the Earth's poles, it will re-orient itself to point in the opposite direction such that its field of view is always minimally obstructed by Earth. In the case of additional Polar satellites similar to SpIRIT, each satellite will independently follow this strategy such that every satellite is co-pointing with one of the HERMES - TP/SP triplets at any given point in time.
            
        \end{itemize}
    
        Note that for the actual HERMES - TP/SP constellation in orbit (or for any constellation of Polar-orbiting satellites) the satellites will drift over time in absence of active orbital control maneuvers, hence the pointing strategy of the satellite triplets would have to be constantly updated and optimised based on the actual positions of the satellites \citep{Sanna_2020}. However, in this work we neglect satellite drift, using instead the simplified configuration with constant relative orbital phases as outlined above.

    \subsection{Simulation Method}
    \label{sec:Simulation Tools}
    
        We model the orbits of each satellite as defined in Section \ref{sec:Satellite Orbits}. We ignore the Moon in our simulations since it occupies such a small fraction ($\sim 0.25\degree$ radius) of the sky, and likewise the influence of the Sun is not relevant as our satellites point at a $90\degree$ offset from the Sun and their FoV only has a $80\degree$ radius. 
        However, we do track the position of Earth for LoS obstruction calculations; from a 550km LEO, Earth has an angular diameter of $\sim 134\degree$, meaning that Earth will maximally encroach on one side of the satellite's FOV by $\sim 47\degree$ when Earth is perpendicular to the satellite's pointing direction (assuming the HERMES instrument's FOV has a radius of $80\degree$).
        We additionally track the operational status of the HERMES instrument on board each satellite based on the particle background maps presented in Section~\ref{sec:Background}.
        
        To compare the localisation capabilities of different sized nano-satellite constellations, we generate a random sample of $10^4$ long and $10^4$ short GRB events for each constellation that we test. The procedure for simulating each GRB is as follows:
        
        \begin{enumerate}
            \item Each GRB is generated with a random right ascension and declination (sampled from a uniform distribution across the celestial sphere), and occurs at a random time in the year.
            
            \item We generate the flux, $F_{GRB}$, of the GRB by selecting a random burst from the Fermi GBM catalog (FERMIGBRST - Fermi GBM Burst Catalog\footnote{https://heasarc.gsfc.nasa.gov/W3Browse/fermi/fermigbrst.html};\citealt{von_Kienlin_2020, Gruber_2014, von_Kienlin_2014, Bhat_2016}). 
            Specifically, we randomly select one of the `Flnc\_Band\_Phtfluxb' values (the average photon flux, in photon/cm2/s between 50 and 300 keV, for a Band function law fit to a single spectrum over the duration of the burst. For long GRBs, we sample from the set of bursts with $T90> 2$ seconds (as reported by the catalog), and for short GRBs we take those with $T90 < 2$ seconds, where $T90$ is defined as the time interval over which the GRB emits 90\% of the total energy of its prompt emission phase.
            
            \item If the GRB occurs within the FOV of one of the co-pointing groups in the constellation, we check whether each satellite has LOS to the coordinates of the GRB or whether they are occulted by the Earth.
            
            For each satellite that has LOS to the burst we calculate the observed GRB flux, $F_{obs}$, based on the incident angle of the GRB to the detector (Equation \ref{eq:HERMES instrument response}). Note that we do not include Poisson noise on the observed flux of the GRB.
           
            We model the flux sensitivity of the HERMES instrument as four times less sensitive than Fermi GBM (as per Section \ref{sec:HERMES - TP/SP}). Given we are sampling `Flnc\_Band\_Phtfluxb' values from the Fermi GBM database, this corresponds to:
            
            \begin{equation}
                F_{\text{lim}} =
                \begin{cases}
                    0.463\text{\,ph}/\text{cm}^2/\text{s} &, \text{Long GRBs} \\
                    1.861\text{\,ph}/\text{cm}^2/\text{s} &, \text{Short GRBs}
                \end{cases}
            \end{equation}
            
            We consider a satellite to detect a GRB if it has LOS to the burst and $F_{\text{obs}} > F_{\text{lim}}$.

            Note that in the simplified context of our simulation the value of $F_{obs}$ is the same for all co-pointing satellites in the constellation, since we model them as pointing at the exact same coordinates. In reality this issue is far more complex - the pointing of each satellite will drift slightly, the response function of each HERMES instrument may differ slightly in shape or absolute value, and there will be incident-angle dependent energy dispersion which scatters high energy photons down to lower energies, influencing the number of photons in the chosen energy band.

            \item We compute the standard deviation on cross-correlation uncertainty $\sigma_{cc}$ according to the observed flux $F_{obs}$ using the power-law fits plus Gaussian scatter model detailed in Table \ref{tab:power-law fits}.
        \end{enumerate}
        
        We note that this method contains several large simplifications regarding the photon energy, noise treatment, and incident angle response of the HERMES instrument and the relation between GRB flux and cross-correlation uncertainty, both of which will impact the accuracy of the results presented in this work. However, as the purpose of this work is to demonstrate the improved localisation performance gained by launching a satellite into a different orbital plane, we proceed with these simplifying assumptions and highlight that our results should only be considered an approximate estimate of the nano-satellite constellation's GRB localisation capabilities to within a factor $\approx 2$ of precision. 

    \subsubsection{Subsample of "observable" GRBs}
        By construction, the method presented above includes the generation of GRB events that cannot be observed by the nano-satellite constellation (either because they are too faint or occur outside the FOV of the constellation), or that are otherwise unsuitable for localisation if too few satellites detect the burst. For these reasons, and to aid in the clarity of our results, we define the following additional conditions that must be met in order to include a GRB in our sample of $10^4$ events that we use as base for our analysis:
        
        \begin{itemize}
            \item The GRB must occur within the FOV of one of the co-pointing groups in the constellation:
            
            \begin{equation*}
                \text{Burst incident angle } \theta \leq 80\degree.
            \end{equation*}
            
            Since the nano-satellite constellation is divided into two co-pointing groups each pointing in opposite directions the burst can only ever be detected by one co-pointing group at a time.
            
            \item The observed flux of the GRB must be higher than the nominal flux limit of the HERMES instrument:
            
            \begin{equation*}
                F_{\text{obs}} = F_{\text{GRB}} \times \cos{\theta } \geq F_{\text{lim}}.
            \end{equation*}
            
            \item At least three satellites in the constellation must directly detect the burst:
            
            \begin{equation*}
                N_{\text{det}} \geq 3.
            \end{equation*}
            
            In order to localise the burst using triangulation techniques at least three independent detections are needed (see Section \ref{subsec:Triangulating GRBs}). Note that we do not enforce that every satellite in the co-pointing group  detects the GRB, since Earth may reduce the effective field of view of some elements, and a non-detection in this context can still contribute to the burst localisation (see Section \ref{subsubsec:Simulating Triangulation} for further discussion). 
            
            \item Every satellite in the co-pointing group must be operational at the time of the burst (i.e., not travelling through a region of high particle flux). We impose this restriction because the GRB localisation capabilities of an n-satellite constellation with one satellite missing are approximately equivalent to localisation statistics reported by an (n-1) satellite constellation (not taking into account the FOV and orbital position of each satellite). Therefore, to avoid such overlap in our results we enforce that every co-pointing satellite is operational at the time of the burst.
            
            \item In the case of the SpIRIT + HERMES - TP/SP constellation (i.e., a constellation with only one satellite in Polar orbit) we impose one additional condition that SpIRIT is part of the co-pointing group that detects the GRB. Without this condition, the localisation capabilities presented for the SpIRIT + HERMES - TP/SP constellation would consist both of bursts that are observed by SpIRIT + 3 HERMES satellites, as well as bursts that are observed by only 3 HERMES satellites, the latter of which would be identical to the localisation capabilities presented for the HERMES - TP/SP constellation alone. Therefore we ensure that SpIRIT is part of the co-pointing group that detects the GRB, which is only true for $\sim 50\%$ of bursts (since SpIRIT can only point towards one of the two fields at any given time). Note that we do not enforce that SpIRIT detects the GRB; we only impose that it is co-pointing with other satellites that detect the GRB.
            
        \end{itemize}
        
        If a simulated GRB meets these conditions, then we include it in our sample and perform the localisation technique described in the following section.
        
    \subsection{Triangulating GRBs}
    \label{subsec:Triangulating GRBs}
        
        \subsubsection{Triangulation Method}
        \label{subsubsec:Triangulation Method}
        
            In this work we follow the localisation method used in \citet{Sanna_2020}, which utilises the arrival time difference of photons at each detector in order to triangulate the position of the GRB. This technique has been used for decades by the IPN to triangulate the positions of GRBs (see e.g., \citealt{Hurley_1999, Hurley_2013}), and as such is appropriate for our estimation purposes, despite some limitations discussed more in Section \ref{subsec:Signal Cross-Correlation Uncertainty}. Here, we provide a brief outline of the method, referring the reader to the work referenced above for full details. 
            
            As GRBs occur at cosmological distances, one can represent the gamma ray photons from the burst as a narrow plane wave travelling through space. Given a network of $n$ detectors distributed throughout space (i.e., a satellite constellation), this plane wave will pass over each detector $i$ at a different global time $t_i$ depending on the coordinates of each detector $\vec{r}_i$. 
            
            If the GRB occurred at right ascension $\alpha$ and declination $\delta$, its Cartesian direction can be represented by unit vector $\textbf{\^{d}}$:
            
            \begin{equation}
               \textbf{\^{d}}(\alpha, \delta) = \{ \cos{\alpha}\cos{\delta}, \sin{\alpha}\cos{\delta}, \sin{\delta} \}.
            \end{equation}
            Defining $t_0$ as the time the GRB signal reaches the origin of the chosen reference frame, the arrival time of photons at detector $i$ is then given by
            \begin{equation}
                t_i = t_0 - \frac{\vec{r_i} \cdot \textbf{\^{d}}}{c},
            \end{equation}
            Where $\vec{r_i}$ represents the position vector of the $i$-th detector. The time delay between two satellites will then be:
            \begin{equation}\label{eq:true time difference}
                \Delta t_{ij}(\textbf{\^{d}}) = \frac{(\vec{r_j} - \vec{r_i})\cdot \textbf{\^{d}}}{c}.
            \end{equation}
            
            In Equation \ref{eq:true time difference}, $\Delta t_{ij}$ represents the difference in arrival time of the signal at each instrument due purely to satellite geometry. The measured value of time delay $\Delta \tau_{ij}$ will depend on this intrinsic time delay plus some offset due to measurement uncertainty:
            
            \begin{equation}
                \Delta \tau_{ij}(\textbf{\^{d}}_{GRB}) = \Delta t_{ij}(\textbf{\^{d}}_{GRB}) + \mathcal{N}(0,\sigma_{cc}),
            \end{equation}
            
            where the term $\mathcal{N}(0,\sigma_{cc})$ represents the uncertainty on the measurement of the arrival time difference. This timing uncertainty is discussed in detail in Sections \ref{subsec:Signal Cross-Correlation Uncertainty} and \ref{subsubsec:Simulating Triangulation}. 
            
            To estimate the true position of the GRB, we compare the measured time-delay pairs $\Delta \tau_{ij}(\textbf{\^{d}}_{GRB})$ from each satellite that detected the GRB to the time delays that would be measured if the GRB were coming from some guess direction, $\Delta t_{ij}(\textbf{\^{d}}_{guess})$. In this work we perform this comparison using a non-linear least squares method: we define the $\chi^2(\textbf{\^{d}}_{guess})$ function as the sum of the squares of the differences between the expected and measured time delays divided by the uncertainty on the signal alignment after cross-correlation:
            
            \begin{multline}\label{eq:Chi Squared Minimisation}
                \chi^2(\textbf{\^{d}}_{guess}) = \\ \sum_{i=0}^{n-2}\sum_{j=i+1}^{n-1} \frac{(\Delta \tau_{ij}(\textbf{\^{d}}_{GRB}) - \Delta t_{ij}(\textbf{\^{d}}_{guess})^2}{\sigma_{\text{cc}}^2}.
            \end{multline}
            
            The coordinates of the GRB can then be estimated by finding the value of $\textbf{\^{d}}_{guess}$ which minimise the $\chi^2$ function. Equation \ref{eq:Chi Squared Minimisation} takes the form of a `minimum $\chi^2$' function with two degrees of freedom (right ascension and declination), and so we estimate the $68\%$ and $90\%$ confidence region of the source localisation as the set of all values of $\alpha, \delta$ where \citep{Avni_1976}:
            
            \begin{equation}\label{eq:Confidence Region}
                \chi^2(\alpha, \delta) - \chi^2_{\text{min, sky}} <=
                \begin{cases}
                    2.3&, 68\%\\
                    4.61&, 90\%
                \end{cases}
            \end{equation}

            Note that Equation \ref{eq:Confidence Region} is a linear approximation for determining the $68\%$ and $90\%$ confidence intervals. In fact this equation is only asymptotically correct for linear models, and as triangulation is a non-linear problem, the degrees of freedom are not necessarily exactly 2. In fact, the exact value of the degrees of freedom depends on data quality and predictiveness of the fit parameters (e.g., \citealt{Janson_2013}).
            
            Finally, we consider the sky coverage and LOS of each satellite in the co-pointing group, which influences GRB localisation depending on whether each satellite directly detects the GRB:
        
            \begin{itemize}
                \item If a satellite detects the GRB, we exclude all directions $\textbf{\^{d}}_{guess}$ from the localisation region that are either outside that satellites FOV or obstructed by the Earth from that satellite's perspective.
                \item If one satellite in a co-pointing group does not detect the GRB, but other satellites in that co-pointing group detect the burst, then we only consider possible burst directions $\textbf{\^{d}}_{guess}$ that were obstructed by Earth from the non-detecting satellite's perspective. 
            \end{itemize}
    
        \subsubsection{Signal Cross-Correlation Uncertainty}
        \label{subsec:Signal Cross-Correlation Uncertainty}
             
            The triangulation method for GRB localisation relies heavily on an accurate estimation of $\Delta \tau_{ij}$ - the measured time difference between the signal arriving at each detector in the network. Classically, this time difference has been computed by binning the light curves observed by each detector with some temporal resolution (which usually depends on the flux of the GRB - brighter bursts can be binned with finer resolution), and then using a cross-correlation algorithm to determine the time offset between the two observed signals \citep{Hurley_1999, Hurley_2013, Sanna_2020}. Note that Equation \ref{eq:Chi Squared Minimisation} therefore presumes that the only source of timing uncertainty is in the signal cross-correlation, which is a valid approximation in our case as the uncertainty on cross-correlation is of the order $\sim 1$\,ms \citep{Sanna_2020} whereas the positional uncertainty on the satellite is $< 30$ m (translating to a temporal accuracy of $< 20$\,ns; \citealt{Fiore_2020}) and the detector has an absolute timing accuracy lower than $0.4\,\mu$s \citep{Evangelista_2020a}. 
            
            \begin{figure}[t!]
            \centering
            \includegraphics[width = \textwidth]{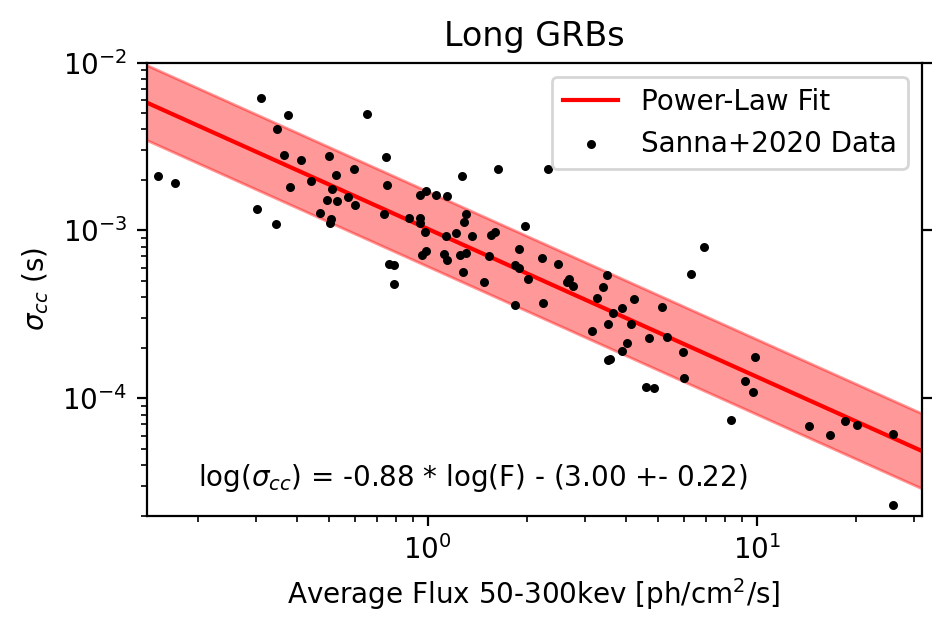}
            \\
            \includegraphics[width = \textwidth]{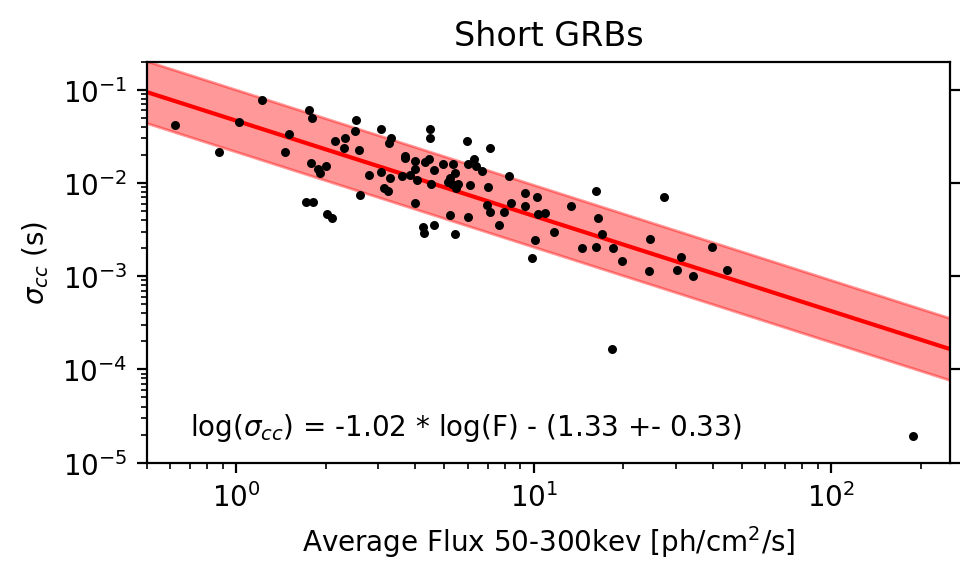}
            \caption{Black points: Standard deviation in the cross-correlation uncertainty ($\sigma_{\text{cc}}$) calculated for a sample of 100 long and 100 short GRBs from \citet{Sanna_2020}. Red line: Best-fit power-law to the black data points, where the $1\sigma$ interval is shown in light red.}
            \label{fig:sigma_cc_modelling}
        \end{figure}
            
            It has been shown by \citet{Burgess_2021} that this `classical' triangulation method has limitations, especially for faint bursts when the number of photons per time bin is small. When compared to a full Bayesian treatment of statistical uncertainties, \citet{Burgess_2021} shows that the classic cross-correlation approach results in an overconfidence of the $1\sigma$ uncertainty on time delay by a factor of $\sim 2-3$, which translates directly into an overconfidence on the burst localisation. Nevertheless, given this paper is a continuation of the work performed in \citet{Sanna_2020} and given the computational benefits in our Monte Carlo simulation approach, we use the classic cross-correlation approach, and note that our results may contain a systematic uncertainty of 2-3 times our reported statistical localisations for the faintest bursts.
            
            \begin{table}[b!]
                \caption{Power-law fits to the data presented in \citet{Sanna_2020} used to calculate the uncertainty in signal arrival-time from the average flux of the GRB between 50-300kev.}
                \centering
                \begin{tabular}{ c  c  c}
                    \hline\hline
                     & Line of Best Fit & Standard Deviation\\
                    \hline
                    Long GRBs & $\log \sigma_{cc} = -0.88 \log(F) - 3$  & $\pm 0.22$ \\
                    Short GRBs & $\log \sigma_{cc} = -1.02 \log(F) - 1.33$ & $\pm 0.33$\\
                    
                    \hline\hline
                \end{tabular}
                \label{tab:power-law fits}
            \end{table}

            In order to determine the impact that GRB flux has on time-delay uncertainty, we adopt the work of \citet{Sanna_2020}, where the standard deviation in the cross-correlation uncertainty, $\sigma_{\text{cc}}$ is reported as a function of average GRB flux
            for a random sample of 100 long and 100 short GRBs observed by Fermi GBM. Specifically, in this work we use the `Flnc\_Band\_Phtfluxb' data type from the Fermi GBM catalog \citep{Gruber_2014, von_Kienlin_2014, Bhat_2016, von_Kienlin_2020}, which is a quantity related to the burst photon flux between 50-300keV. The results from \citet{Sanna_2020} are reported here in Figure \ref{fig:sigma_cc_modelling} as black data points. In this figure, each data point represents the results from 1000 Monte-Carlo (MC) simulations which calculated the difference between the true arrival time delay and the value inferred from cross-correlation methods. The resulting distribution of time-offset values is then fit with a Gaussian function with a mean of zero and a standard deviation $\sigma_{cc}$. To perform 1000 MC trials for each burst, the GRB light curve observed by Fermi GBM is used as a template from which unique GRB light curves can be generated by both rescaling the detector effective area to match that of the HERMES instrument and applying a Poissonian randomisation of the counts contained in each bin of the template.

        To generalise the results of \citet{Sanna_2020} we construct a power-law fit to this data (Figure \ref{fig:sigma_cc_modelling}, red lines), assuming that the residuals follow a Gaussian distribution. The parameters for our fits are given in Table \ref{tab:power-law fits}.
        
        We use these relations to compute the uncertainty on signal cross-correlation, $\sigma_{cc}$, given the observed flux of a burst, which depends not only on the flux of the GRB as seen from LEO, but also on the incident angle of the burst to the detector ($F_{obs} = F_{GRB} \cos \theta$).
        
\begin{figure}[t!]
            \begin{center}
            \includegraphics[width=\textwidth]{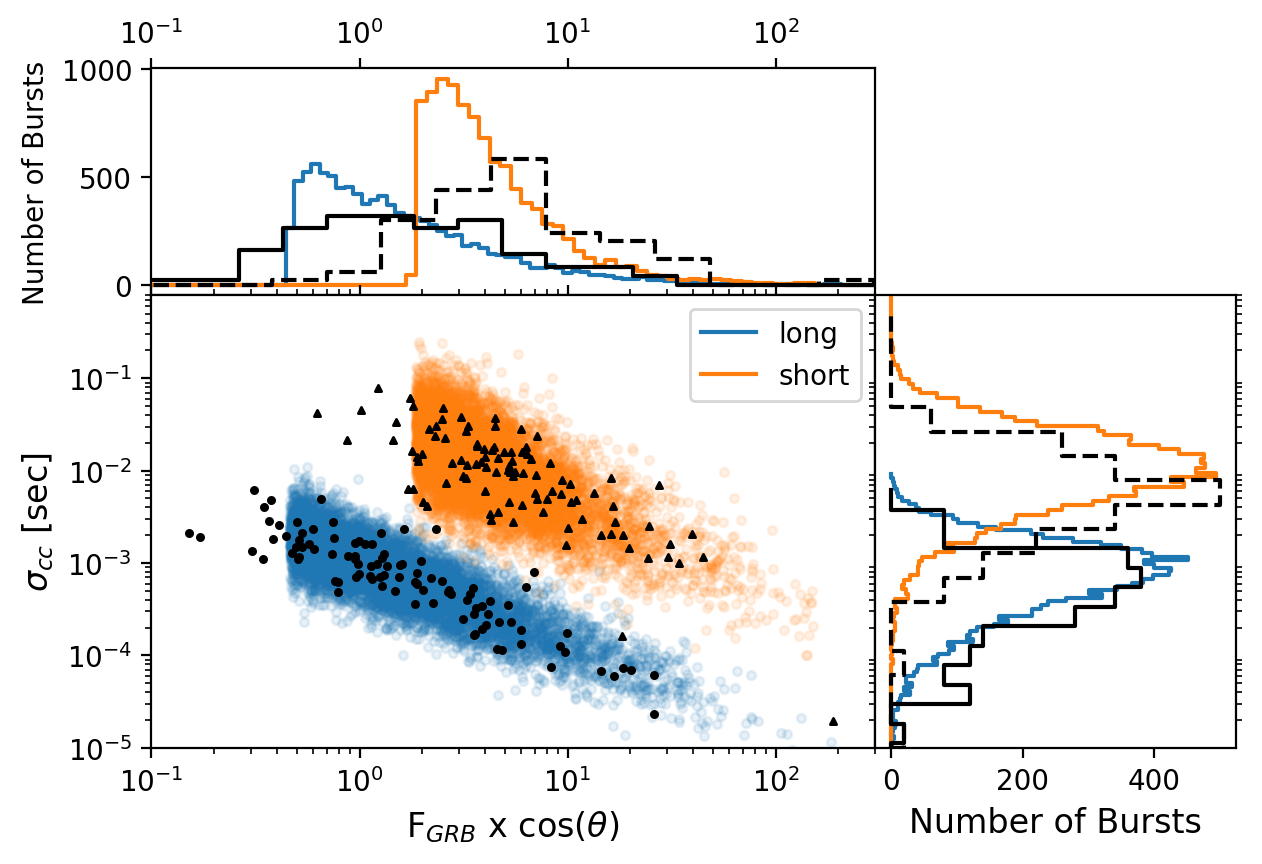}
            \caption{Scatter plot showing the relation between the observed flux of the GRB ($F_{obs} = F_{GRB}\cos\theta $) and the expected standard deviation on timing uncertainty after cross-correlation $\sigma_{cc}$ from $10^4$ MC trials. The top panel plots the distribution of observed average GRB fluxes, and right hand panel show histograms for the resulting distribution of cross-correlation uncertainty as derived from the given flux distributions.
            For comparison, in black we also plot the 100 long GRBs (circles, solid histograms) and 100 short GRBs (triangles, dashed histograms) from \citet{Sanna_2020} which were used as a basis for our model. We have arbitrarily scaled the histograms of the \citet{Sanna_2020} data by eye so that they match the scale of our simulated data.}
            \label{fig:fig:observed flux sigma scatter}
            \end{center}
        \end{figure}

        Figure \ref{fig:fig:observed flux sigma scatter} plots the results from $10^4$ Mote-Carlo trials using this method to calculate the standard deviation on cross-correlation uncertainty. As expected, the satellites in our simulated constellation observe a lower GRB flux on average than those bursts used by \citet{Sanna_2020} (black points/curves), since we take into account the HERMES instruments' approximate cosine dependence on photon incident angle. Otherwise we find good agreement with the results presented in \citet{Sanna_2020}.
        
        Note that this method contains several assumptions, justified by our aim to forecast performance of a future constellation rather than a detailed characterisation of the localisation algorithm applied to actual data. Firstly, we are assuming that average GRB flux is as accurate predictor of cross-correlation uncertainty. For a more realistic simulation, one could follow the same procedure as performed in \citet{Sanna_2020} and perform the cross-correlation of GRB light curves to determine the `measured' arrival time difference. 

        Secondly, our simulation assumes that the uncertainty on signal cross-correlation between two detectors $\sigma_{cc,ij}$ is uncorrelated between pairs of detectors. In reality, we would expect this quantity to have some correlation between pairs of satellites, e.g., $\sigma_{cc,ij}$ should have some relation to $\sigma_{cc,ik}$ since both involve the signal observed by detector $i$. However, understanding these correlations is non-trivial, and beyond the scope of this work.

    \subsubsection{Simulating GRB Localisation}
    \label{subsubsec:Simulating Triangulation}
        
        For each GRB that meets the conditions defined in Section \ref{sec:Simulation Tools} we perform the triangulation procedure described in Section \ref{subsubsec:Triangulation Method}. We start by constructing a set of `measured' time-difference pairs $\Delta \tau_{ij} = \Delta t_{ij}(\textbf{\^{d}}_{GRB}) + \mathcal{N}(0,\sigma_{cc})$ between each satellite which detected the event, where $\mathcal{N}(0,\sigma_{cc})$ represents a random variable sampled from a Gaussian distribution with a mean of zero and standard deviation $\sigma_{cc}$, as defined above. 
        
        Using these values we calculate Equation \ref{eq:Chi Squared Minimisation} for each point on the sky, where we represent the whole sky using a grid of $660048$ points uniformly distributed on the celestial sphere (giving each point an area of $\sim 0.0625$ deg$^2$), with positions generated by the Fibonacci lattice method \citep{Gonzalez_2009}. For computational efficiency, in cases where the localisation is not expected to reach $<1$ deg$^2$ (e.g., when localising short GRBs, or when testing constellations with fewer satellites in Polar orbit), a grid of 41,253 points ($1$ deg$^2$ resolution), or $165012$ points ($0.25$ deg$^2$ resolution) is used.
        
        Finally, we determine the $68\%$ or $90\%$ localisation confidence regions using the values defined in Equation \ref{eq:Confidence Region}, excluding those points that are disallowed by LOS arguments as defined in Section \ref{subsubsec:Triangulation Method}.
        
        We note here that when calculating the time-difference pairs $\Delta \tau_{ij}$ and the timing uncertainty on the signal cross-correlation $\sigma_{cc}$, we presume that the satellites are stationary for the duration of the burst. In reality, the relative motion of each satellite introduces a small difference in the arrival time of photons at each detector. We estimate that as a worst-case scenario - a satellite in LEO moving at $10$km/s in the direction of the burst - would detect the final photon from a typical long GRB (approximately $\sim 10$\,sec) $0.3$\,ms before a stationary satellite, meaning that the moving satellite would effectively observe the signal of the GRB  `compressed' in time by $0.3$\,ms, which is comparable to the cross-correlation accuracy achieved for the brightest $\sim 20$\% of long GRBs. However, given the position and velocity of each satellite in the SpIRIT + HERMES constellation are known to a high accuracy, in principle this effect is entirely deterministic and can be accounted for in a sufficiently sophisticated cross-correlation algorithm. For this reason we choose to neglect the relative motion of each satellite in this work, leaving its inclusion to a future simulation of the HERMES constellation's localisation capabilities, and note that this assumption may introduce some systematic uncertainty into our results for the best localised long GRBs.
        
\section{Results and Discussion}
\label{sec:Results and Discussion}

    Following the method outlined in Section \ref{sec:Simulation Tools}, we simulate the localisation capabilities of HERMES - TP/SP, as well as SpIRIT + HERMES - TP/SP and constellations featuring additional SpIRIT-like satellites in Polar orbit. The following sections present the localisation results for a sample of $10^4$ long and $10^4$ short GRBs.

    \subsection{Constellation Sky Coverage}
    \label{subsec: Constellation Sky Coverage}
        
        Figure \ref{fig:Constellation Pointing & Visibility} demonstrates that the combined SpIRIT + HERMES - TP/SP constellation is only able to detect a maximum of $60\%$ of long GRBs and $\sim 50\%$ of short GRBs, which is to be expected since the FWHM of the HERMES instrument is $60\degree$ meaning that each FOV covers $\sim 25\%$ of the sky at FWHM, and the constellation is divided into two co-pointing groups. We note that the constellation is able to detect a higher fraction of long GRBs than short GRBs because of the different limits we have placed on the average flux sensitivity for the two types of burst (see Section \ref{sec:Simulation Tools}).
        
        We find that for a constellation with a single satellite in polar orbit, approximately $\sim 10\%$ of bursts can be detected by four orbital elements simultaneously (SpIRIT + all three satellites in one of the HERMES triplets). This low fraction is both due to the transit of each orbital element through high particle flux regions (e.g., SpIRIT over the poles and through the SAA; HERMES through the edge of the SAA), as well as in part of each satellite's FOV being obstructed by the Earth at the point in its orbit when Earth is perpendicular to its LOS.
        
        For a constellation of 4 Polar satellites + HERMES - TP/SP we find similarly that only $\sim 10\%$ of GRBs can be detected by the all 5 of the co-pointing satellites (2 Polar satellites + 3 HERMES - TP/SP satellites), while $\sim 25\%$ of bursts are detected by four satellites (1 Polar satellite + 3 HERMES - TP/SP, or two Polar satellites + 2 HERMES - TP/SP), with one satellite missing the observation due to crossing a high particle flux region or LOS obstruction. The case is similar for six Polar satellites, where $\lesssim 10\%$ of bursts are detected by all of the co-pointing satellites due to the intermittent passage of each orbital element through a region of high particle flux.
        
        In the case of launching additional Polar satellites, the benefit comes from the fact that each burst is detected by a larger number of satellites on average. For example, in the case of a single Polar satellite, only $10\%$ of bursts are detected by four satellites, while with six Polar satellites $\sim 40\%$ of GRBs are detected by four or more satellites, enabling more accurate burst localisations on average despite the fact that not all Polar satellites are observing the event.
    
        \begin{figure}[t!]
            \begin{center}
            \includegraphics[width=\textwidth]{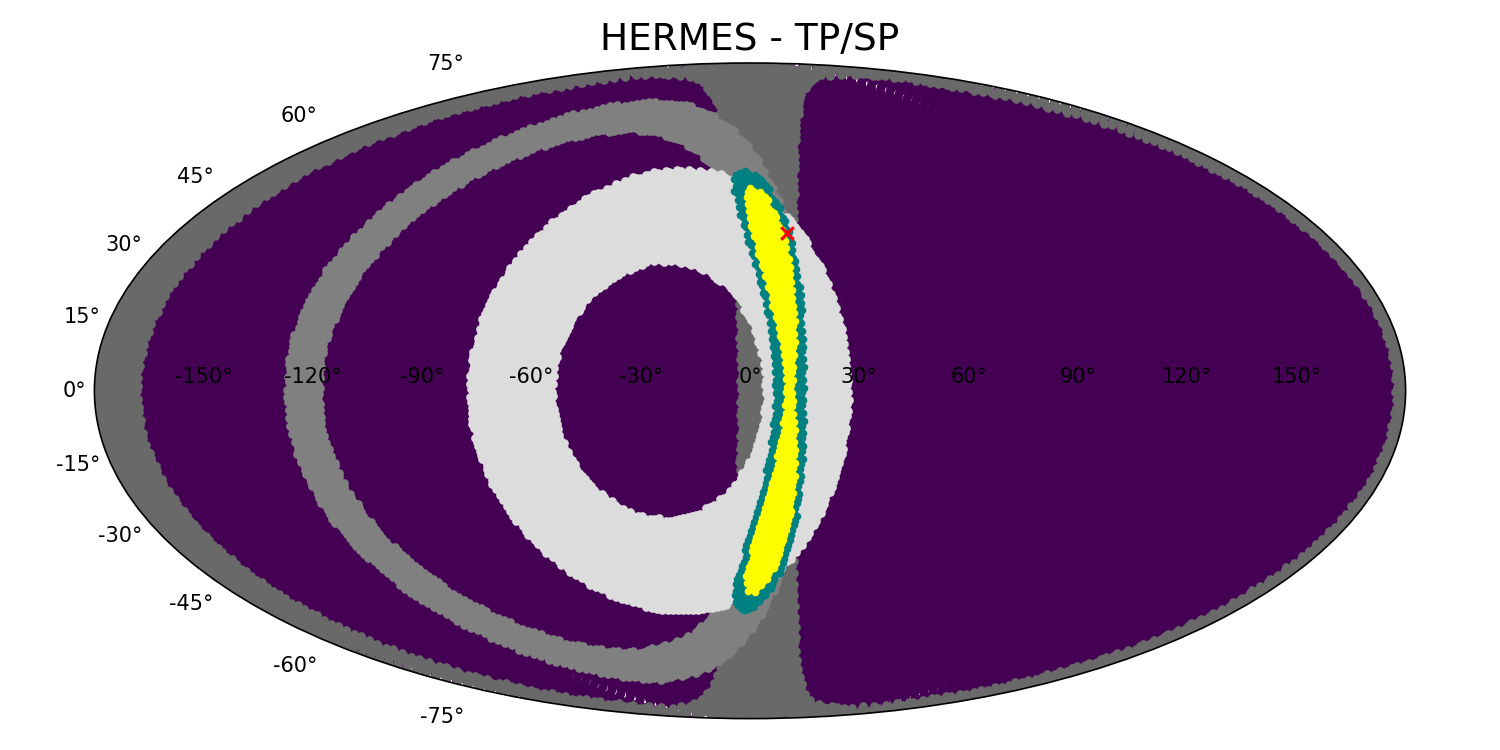}\\
            \includegraphics[width=\textwidth]{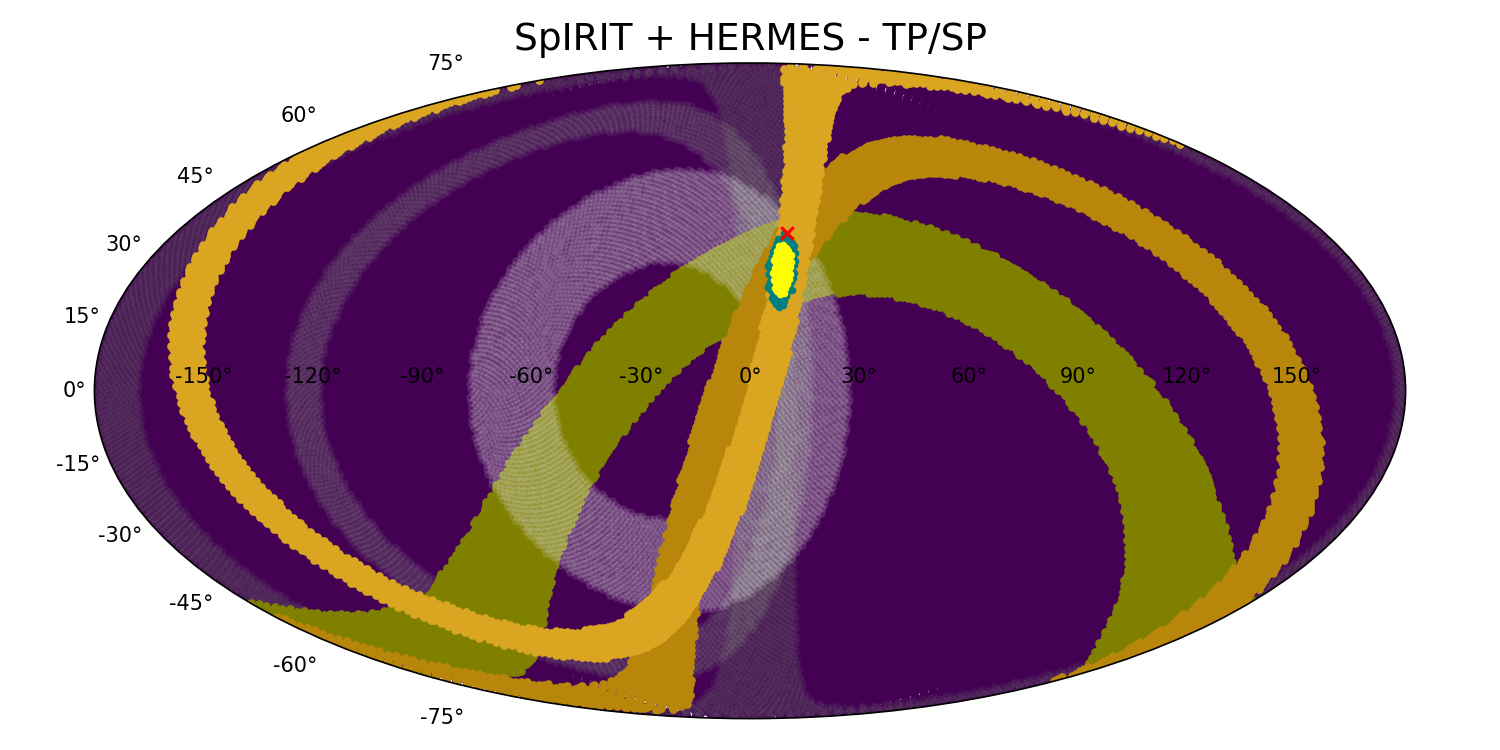}
                \caption{Comparison of the $68\%$ (yellow) and $90\%$ (teal) confidence regions for the location of a long GRB as measured by the HERMES - TP/SP and by the SpIRIT + HERMES - TP/SP constellations for one realisation in our Monte Carlo simulation. The red cross denotes the location of the GRB on the sky. The grey lines represent the $1\sigma$ error annuli between each pair of satellites in the HERMES constellation. The gold lines represent the error annuli for the three additional satellite pairs that can be made when SpIRIT is included in the HERMES - TP/SP.}
            \label{fig:Localisation Region Comparison}
            \end{center}
        \end{figure}
    
    \subsection{SpIRIT + HERMES - TP/SP Localisation}
    \label{subsec:SpIRIT + HERMES - TP/SP Localisation}
        
        The primary advantage of augmenting the HERMES-TP/SP constellation with the SpIRIT satellite is that the additional detector in Polar orbit greatly reduces the localisation degeneracy in declination when SpIRIT is located at high latitudes. 
        
        Because each satellite in the HERMES - TP/SP constellation orbits in the equatorial plane and align their LOS in the same direction, there is limited accuracy and intrinsic degeneracy in the declination coordinate of the burst - for example, it is equally possible that a burst came from declination $\delta = 30\degree$ or $\delta = -30\degree$. However, SpIRIT's Polar orbit gives a large average baseline perpendicular to the orbital plane of the HERMES - TP/SP, which allows for the degeneracy in declination to be broken if SpIRIT is able to detect the event. Indeed, even if SpIRIT doesn't directly detect the event while co-pointing with the other observing satellites, a non-detection would still provide useful information to rule out uncertainty regions enclosed in SpIRIT's FoV. An example uncertainty region for HERMES - TP/SP and SpIRIT + HERMES - TP/SP is given in Figure \ref{fig:Localisation Region Comparison}, which demonstrates the increased localisation accuracy in the declination coordinate when including SpIRIT.
        
        It should be noted that another way to resolve this degeneracy is by adjusting the pointing direction of each satellite so that they do not co-point at the same field. Doing so could allow the constellation to exploit the HERMES instruments' cosine dependence on photon incident angle to localise bursts by taking into account the ratio of GRB fluence at each detector. However, this would lead to a trade-off between GRB localisation accuracy and number of GRBs localised, since a smaller fraction of the sky would be observed by all three satellites. This trade-off is something that will be explored in future work.
        
\begin{figure}[t!]
            \centering
            \includegraphics[width=\columnwidth]{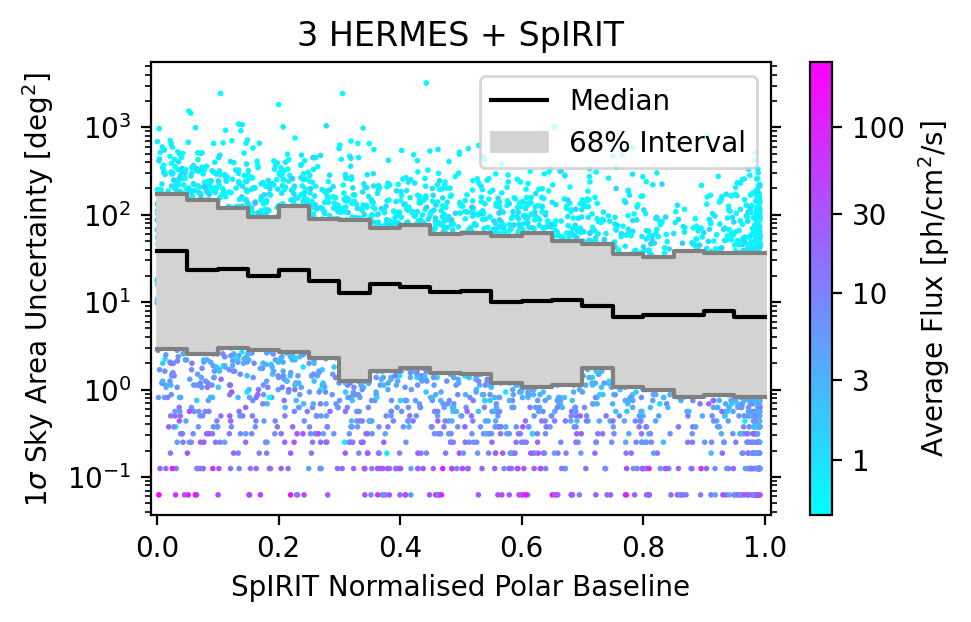}
            \caption{Plot of the $1\sigma$ localisation region for $\sim 10^4$ simulated long GRB events observed by the SpIRIT + HERMES - TP/SP constellation. The x-axis shows SpIRIT's baseline perpendicular to the equatorial plane for each burst normalised by its maximum baseline (maximum baseline = Earth radius + 550km orbital altitude). Note here that `baseline' refers to SpIRIT's distance perpendicular to the equatorial plane. The black line shows the median of the data points within the bin (binned into intervals of 0.05 on normalised polar baseline), and the grey region contains the central $68\%$ of data in each bin. Individual data points are shown for outliers, with the colour denoting the average flux of the GRB.}
            \label{fig:HERMES+SpIRIT SpIRIT z-proj}
        \end{figure}

        Figure \ref{fig:HERMES+SpIRIT SpIRIT z-proj} plots the area of the $1\sigma$ localisation region for long GRBs detected by SpIRIT + 3 HERMES - TP/SP satellites against SpIRIT's distance from the equatorial plane, where we have colour-coded each data point by the average flux of the GRB. While the localisation accuracy depends strongly on the flux of the GRB, we find that SpIRIT's polar orbit is able to dramatically improve burst localisations, such that the constellation's capabilities improve by a factor of $\sim 5$ when SpIRIT is over the poles compared to when it is at the equator. When SpIRIT is close to its maximum distance from the equatorial plane, we expect to achieve statistical burst localisations within $\sim 10$ deg$^2$ for $50\%$ of long GRBs, and localisations within a few square degrees for bursts with an observed time-averaged flux higher than $\sim 5$ph/cm$^2$/s ($\sim 15\%$ of long GRBs). 
        
        Note that while in the context of our simulation the constellation achieves localisations $<1$\,deg$^2$ for very bright bursts when SpIRIT is near the poles, quantifying with precision such cases goes beyond the simplifying assumptions upon which our framework is based.

        In the following analysis we divide our results into long and short GRBs.
        
        \subsubsection{Long GRBs}
        \label{subsection:HERMES + SpIRIT Long}
        
\begin{figure}[b!]
                \centering
                \includegraphics[width=0.94\columnwidth]{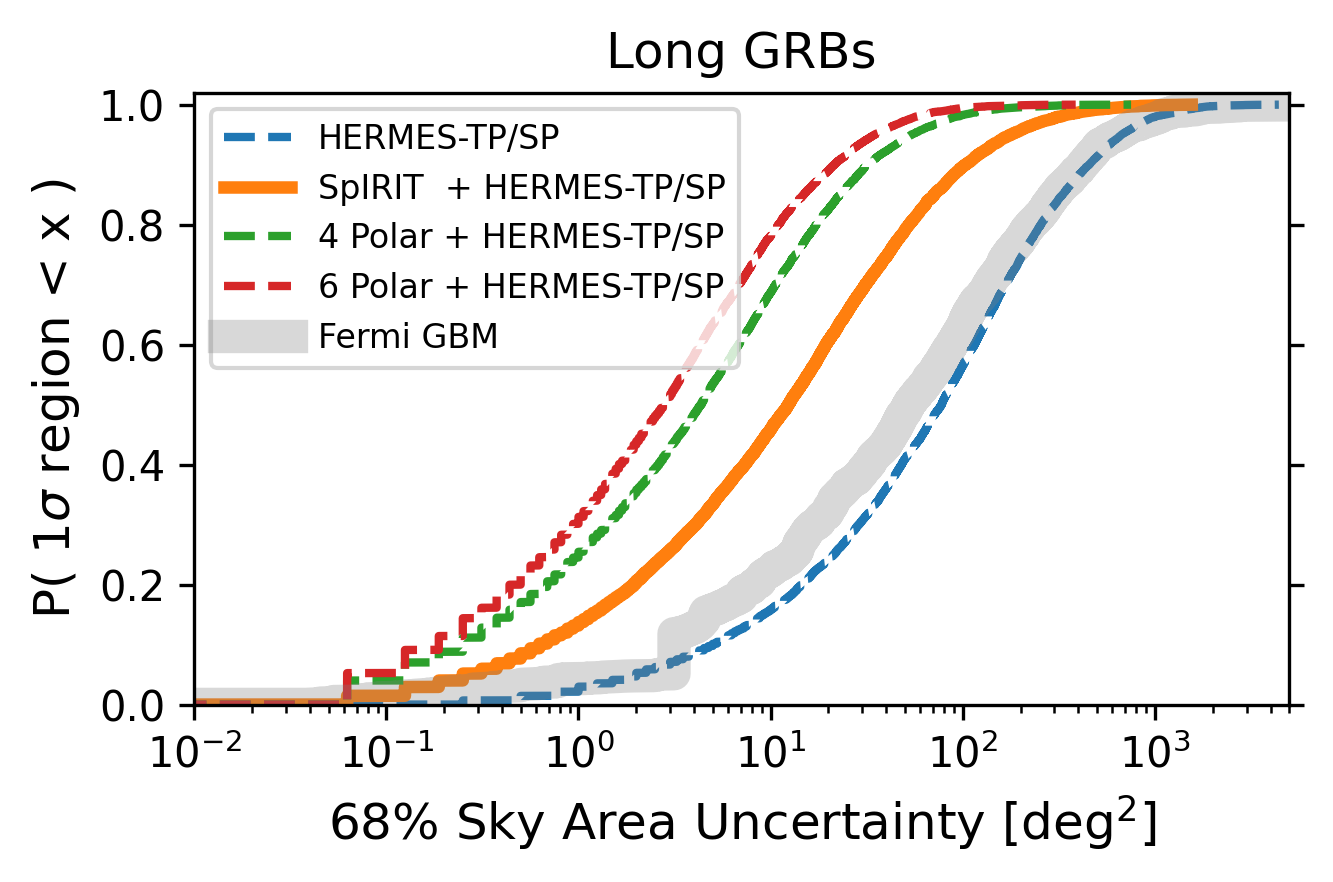}
                \caption{Cumulative probability of localising a long GRB within a given $1\sigma$ region on the sky for $n$ SpIRIT-like satellites in Polar orbit + HERMES - TP/SP. The $n_\text{polar} = 0$ constellation (blue line) represents the HERMES - TP/SP constellation operating alone.}
                \label{fig:Constellation CDF long}
            \end{figure}

            Figure \ref{fig:Constellation CDF long} compares the cumulative distributions of the $1\sigma$ localisation region for $10^4$ long GRBs localised by HERMES - TP/SP and SpIRIT + HERMES - TP/SP. For comparison we also plot the $68\%$ statistical sky area uncertainties reported in the fourth Fermi/GBM catalog \citep{von_Kienlin_2020}. Note here that we are comparing the statistical GRB localisation capabilities only - while both instruments also have an associated systematic uncertainty, we cannot estimate these for the nano-satellite constellation since to do so would require in-flight data, which is impossible since the constellation has not been launched yet. For reference, the systematic uncertainty for Fermi GBM is approximately $\sim 2-4\degree$ \citep{Goldstein_2020}, and it will likely be comparable to this for the nano-satellite constellation.
            
            \begin{figure*}[hbt!]
            \centering
            \begin{subfigure}[b]{0.45\linewidth}
                \centering
                \includegraphics[width = \linewidth]{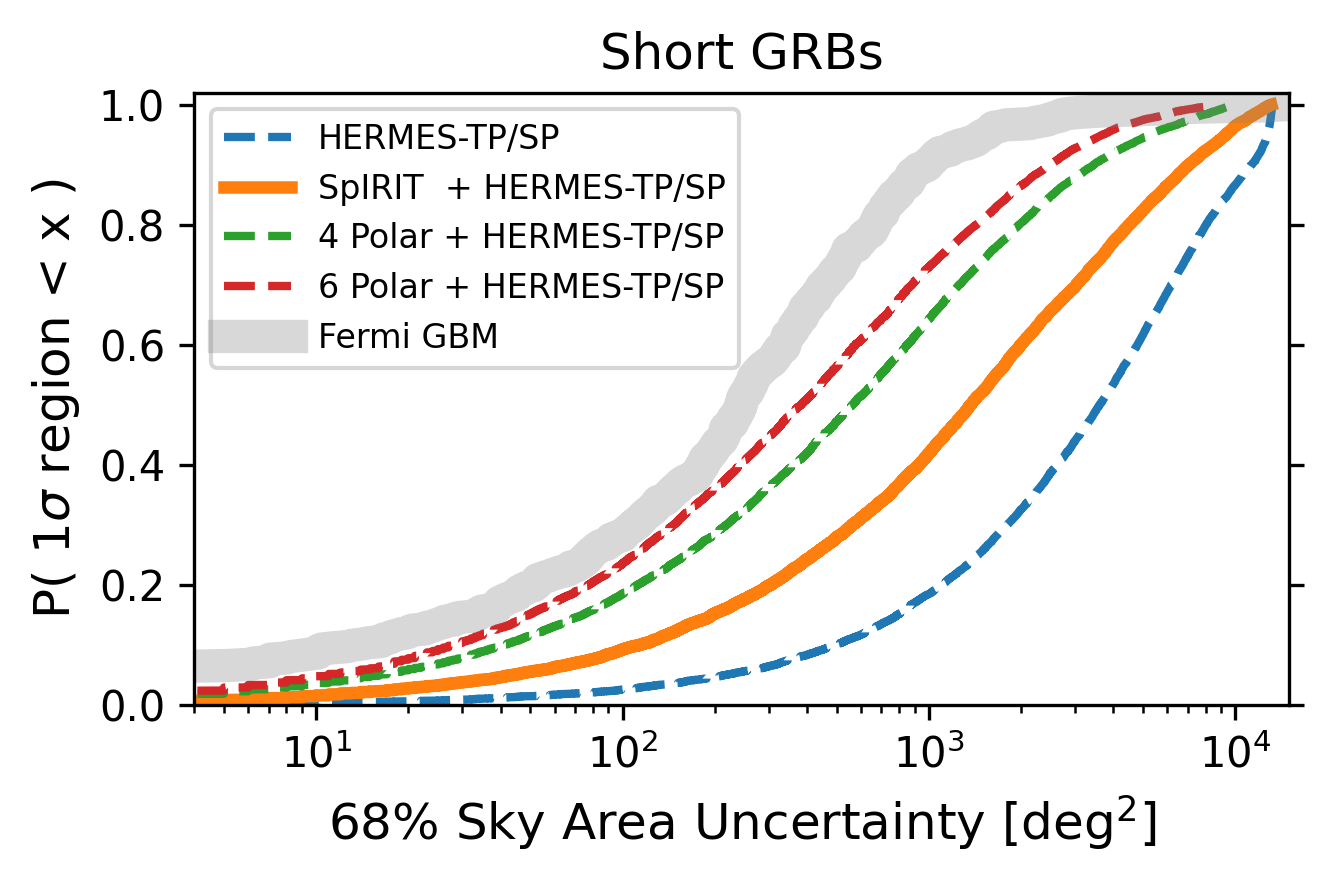}
                \caption{68\% Confidence Interval}
                \label{fig:HERMES+SpIRIT short 68}
            \end{subfigure}
            \begin{subfigure}[b]{0.45\linewidth}
                \centering
                \includegraphics[width = \linewidth]{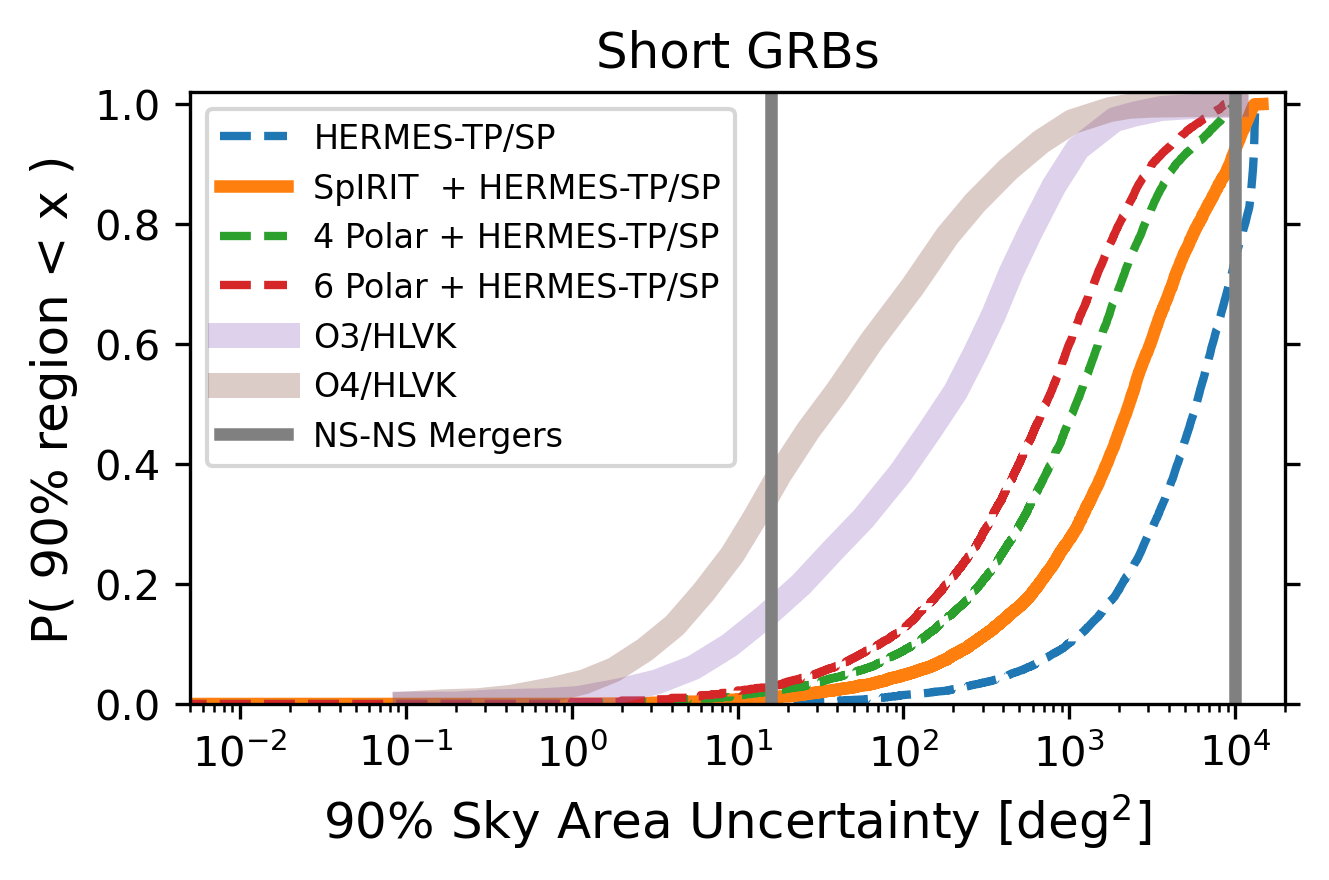}
                \caption{90\% Confidence Interval}
                \label{fig:HERMES+SpIRIT short 90}
            \end{subfigure}
            \caption{Cumulative probability of localising a short GRB within a given area on the sky for $n$ SpIRIT-like satellites in Polar orbit + HERMES - TP/SP. The $n_\text{polar} = 0$ constellation (blue line) represents the HERMES - TP/SP constellation operating alone. \textit{Left}: 68\% confidence intervals. \textit{Right}: 90\% confidence intervals. The dashed lines represent the highest expected localisation capabilities of the advanced LIGO Hanford and Livingstone, advanced VIRGO and KAGRA GW detectors to localise BNS coalescences in O3 and O4 respectively (note that for O3, detector sensitivities were taken to be representative of the first 3 months of observations for aLIGO Hanford and Livingston, and AdV, and the highest expected O3 sensitivity for KAGRA). See \citep{Abbott_2020} Figure 6 for details). The two vertical grey lines represent the prompt reported 90\% confidence regions for the two binary neutron star coalescences observed to date; GW170817, and GW190425.}
            \label{fig:HERMES+SpIRIT short}
        \end{figure*}

            The inclusion of SpIRIT greatly enhances the accuracy of GRB localisations compared to the HERMES - TP/SP alone. SpIRIT + HERMES - TP/SP is able to localise $60\%$ of long GRBs within $\sim 28$ deg$^2$, which corresponds to an uncertainty radius on the sky of $\sim 3\degree$ (presuming the uncertainty region is circular), and $90\%$ of long GRBs within $150$ deg$^2$ ($\sim 7\degree$ radius). In reality the uncertainty regions are irregular in shape, but a circle is a good approximation when the burst is relatively well localised. This represents an improvement by a factor of $\sim 5$ over the HERMES - TP/SP alone, which is only able to localise $\sim 50\%$ of GRBs to within $100$\,deg$^2$ on the sky.
 
            Figure \ref{fig:Constellation CDF long} shows that the combined SpIRIT + HERMES - TP/SP constellation will have statistical GRB localisation capabilities comparable to those reported by Fermi GBM. This result demonstrates the potential cost-effectiveness of using modular nano-satellite astronomy to achieve specific science goals, as a constellation of several relatively inexpensive nano-satellites is expected to achieve a localisation accuracy similar to the substantially more complex and expensive Fermi GBM. In fact, estimating an approximate cost of US \$150M for the Fermi GBM instrument ($\sim 25\%$ of the total cost of the mission (Fermi Gamma-ray Space Telescope Q\&A on the GLAST Mission\footnote{\url{https://www.nasa.gov/mission_pages/GLAST/main/questions_answers.html}})) and of US \$2.5M for each element of the nano-satellite constellation, there is a possible factor 10x in savings using a distributed aperture solution for localising long GRB afterglows. We note however that the disadvantage to using a HERMES-like distributed aperture solution with the current design is sky coverage: the Fermi GBM can observe over half the sky at any time (i.e., the whole sky not obstructed by the Earth \citep{Meegan_2009}), whereas the FoV of a single HERMES instrument only achieves $\sim 25\%$ sky coverage at FWHM, meaning that achieving all-sky coverage with a nano-satellite constellation with overlapped FoV's would partially offset the cost savings (although there would be further gains from mass production of the constellation elements). Furthermore, it is non-trivial to achieve maximal overlap of the FOV of each nano-satellite element, as it requires continual optimisation of the pointing directions of each satellite to account for orbital drift and different orbital configurations \citep{Colagrossi_2020}.

        \subsubsection{Short GRBs}
        \label{sec:HERMES+SpIRIT short}
            
            Figure \ref{fig:HERMES+SpIRIT short} compares the localisation capabilities of the HERMES - TP/SP and combined SpIRIT + HERMES - TP/SP constellations to both the Fermi GBM (where we have plot the Fermi localisation uncertainty using same method as discussed in Section \ref{subsection:HERMES + SpIRIT Long}) and to the theoretical current and future localisation capabilities of the Advanced LIGO Hanford and Livingston, Advanced VIRGO and KAGRA gravitational wave detectors (\citet{Abbott_2020}; $90\%$ confidence regions, Figure \ref{fig:HERMES+SpIRIT short 90}). Note that this reflects the predicted performance of the four gravitational wave detectors operating concurrently during O3 and O4 - in reality KAGRA did not participate in O3, but is planned for joint operation in O4. We have also included the reported prompt localisations for both of the BNS coalescences observed by GW detectors; GW170817 \citep{Abbott_2020} and GW190425 \citep{Abbott_2021}. While the error box for GW190425 was later refined, here we compare the prompt localisation capability to ensure a fair comparison to a nano-satellite constellation. Note that the HERMES - TP/SP constellation is less effective at localising short GRBs compared to long GRBs due to the increased difficulty in precise signal cross-correlation for short GRBs, although a future instrument redesign with increased collecting area would mitigate this current limitation \citep{Sanna_2020}. 
            
            The addition of SpIRIT to HERMES - TP/SP is able to substantially improve the localisation of short GRBs. HERMES - TP/SP alone has a 90\% $1\sigma$ localisation range of $\sim 215-13000$ deg$^2$, while the SpIRIT + HERMES - TP/SP constellation's 90\% $1\sigma$ range is $\sim 50-8450$ deg$^2$, where we have defined the 90\% $1\sigma$ localisation interval as the range of $1\sigma$ confidence regions spanning 5\% to 95\% of the cumulative distribution.
            
            Despite this improvement in the constellation's ability to localise short GRBs, in terms of multi-messenger observations of short GRBs/BNS coalescences the combined SpIRIT + HERMES-TP/SP constellation is not expected to dramatically improve upon existing capabilities to localise these sources. Not only is the Fermi GBM more effective by a factor of $\sim 5-10\times$ at the EM localisation of short GRBs, but current-generation GW instruments are theoretically capable of localising $> 50\%$ of short GRBs within $\sim 200$ deg$^2$ (and within $\sim 40$ deg$^2$ in the future O4 observing runs) from the GW signal alone \citep{Abbott_2020}, which exceeds the localisation capabilities of the nano-satellite constellation. Therefore, in the domain of multi-messenger astronomy the utility of the SpIRIT + HERMES - TP/SP constellation will be in the constellation's sky coverage ($\sim 50\%$) and flux sensitivity, which gives it a high probability of detecting the electromagnetic counterpart to a compact binary event observed by GW detectors. Such dual detections are critically needed to improve our understanding of these compact binary mergers, as currently only one simultaneous detection of the GW signal from a BNS merger (GW170817) and a short GRB (GRB170817A) has been recorded to date \citep{Abbott_2017}.
            
            \begin{figure*}[hbt!]
                \centering
                \begin{subfigure}[b]{0.45\linewidth}
                    \centering
                    \includegraphics[width = \linewidth]{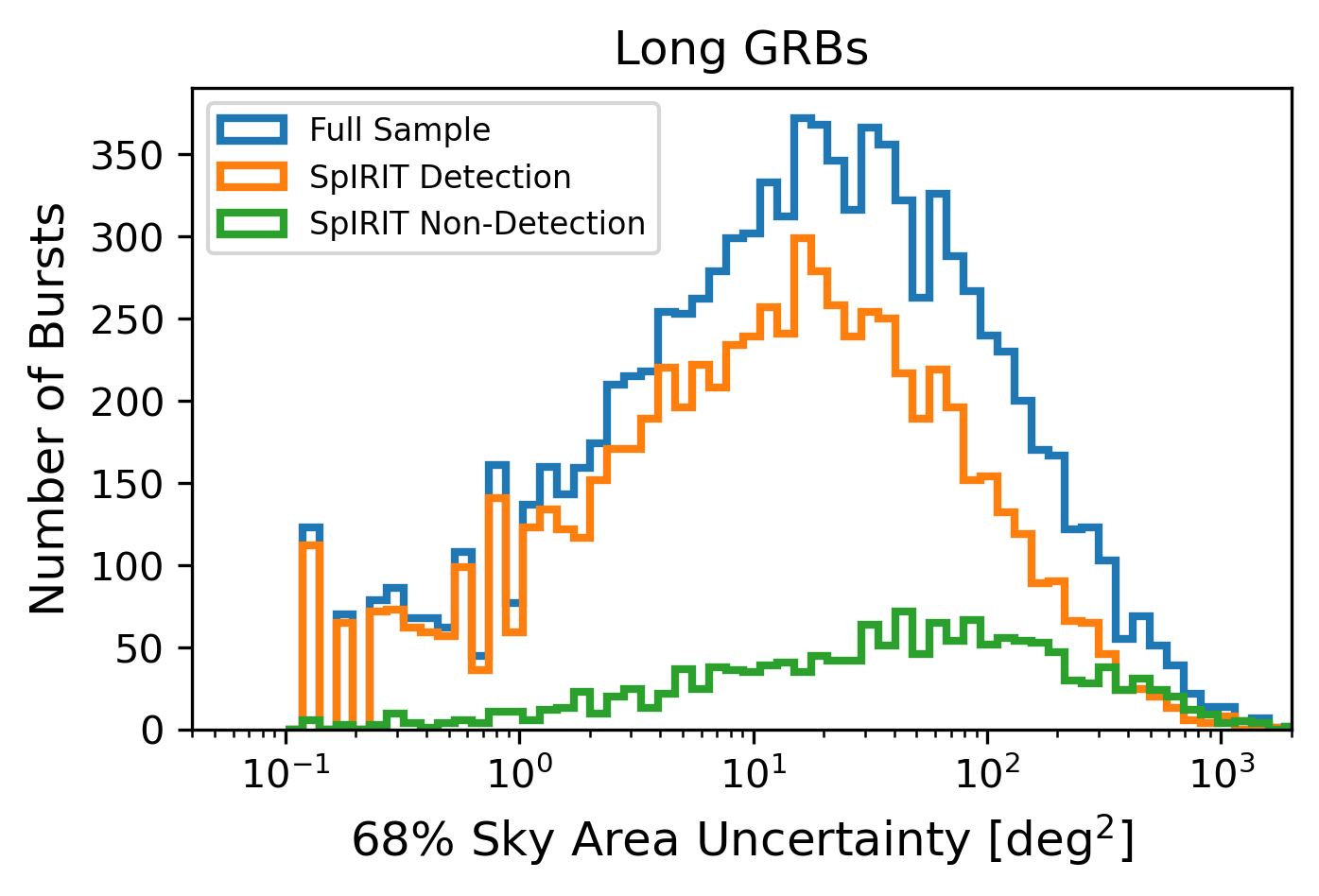}
                    \label{fig:sample_analysis_long}
                \end{subfigure}
                \begin{subfigure}[b]{0.45\linewidth}
                    \centering
                    \includegraphics[width = \linewidth]{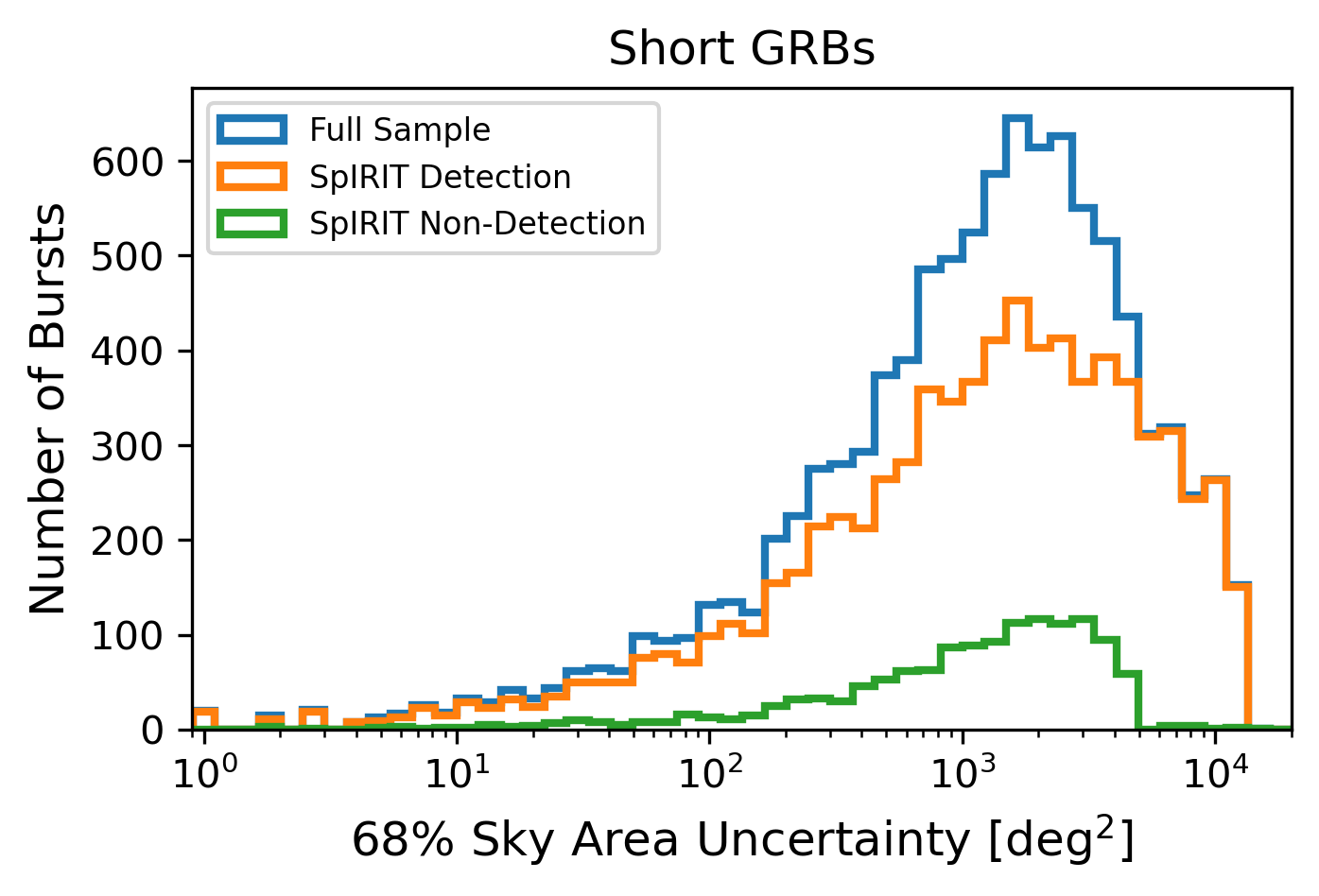}
                    \label{fig:sample_analysis_short}
                \end{subfigure}
                \caption{Localisation histograms for the sample of $10^4$ GRBs detected by the SpIRIT + HERMES - TP/SP Constellation. The blue curve represents the full sample of simulated GRBs, while the orange and green curves compare the localisation capabilities of the constellation when SpIRIT either detects or does not detect the burst. \textit{Left}: Long GRB sample. \textit{Right}: Short GRB sample.}
                \label{fig:HERMES+SpIRIT Sample Analysis}
            \end{figure*}
   
         \subsubsection{Utilising SpIRIT Non-Detections}
         \label{subsubsec:SpIRIT Non-Detections}
         
            The inclusion of SpIRIT into the HERMES - TP/SP enables improvements to the GRB localisation capabilities of the HERMES - TP/SP both when SpIRIT directly detects the burst, but also when SpIRIT does not directly detect the burst. The improvement in the latter case is a product of SpIRIT's high inclination orbit, since when SpIRIT is over the poles then a considerable section of the HERMES triplet's FOV is obstructed by the Earth from SpIRIT's perspective. This means that a SpIRIT non-detection can  impose relatively strong constraints on possible source location under these conditions, assuming that SpIRIT is co-pointing at the same field as the HERMES triplet that detects the burst, and that the noise level is low enough for SpIRIT that we would expect to detect the GRB signal.
            
            Figure \ref{fig:HERMES+SpIRIT Sample Analysis} compares the size of the $1\sigma$ confidence region for long and short GRBs, where the results have been divided depending on whether SpIRIT detects the GRB or not. Note that in all cases, SpIRIT is co-pointing with the HERMES triplet that detects the GRB. In both cases, we find that SpIRIT makes a direct detection of the GRB for $\sim 85\%$ of bursts in our `observable' sample, which represents the average fraction of the HERMES triplet's FOV that is simultaneously observed by SpIRIT.
            
            In the case of long GRBs we find that the size of the $1\sigma$ localisation region is always smaller when SpIRIT directly detects the burst compared to when SpIRIT does not detect the GRB. This is to be expected, since for HERMES - TP/SP the typical size of the long GRB localisation region is much lower than the sky coverage of the triplet, meaning that a SpIRIT non-detection will only improve the source localisation in the unlikely case that the burst occurs at coordinates close to the edge of Earth as seen from SpIRIT's perspective.
            
            Conversely, in the case of short GRBs we find that in some cases it is possible to achieve more accurate GRB localisations when SpIRIT does not detect the burst. These cases arise for faint short GRBs where the typical uncertainty on signal cross-correlation is large, meaning that the typical $1\sigma$ localisation region is comparable to the sky coverage of the constellation ($\geq 10^4$\,deg$^2$). In such cases, SpIRIT non-detections enable more accurate GRB localisations as they restrict the possible location of the GRB to those regions of sky that are obstructed by the Earth at the time of the burst, which is why there are practically no bursts with $1\sigma$ localisation regions larger than $\sim 5\times10^3$\,deg$^2$ in the green curve in Figure \ref{fig:HERMES+SpIRIT Sample Analysis} (right panel).

    \subsection{Additional Polar Satellites}
    \label{subsec:Additional Polar Satellites}
    
        After demonstrating the benefits of adding one Polar satellite to the HERMES - TP/SP constellation, it is natural to investigate the localisation potential of launching additional satellites into Polar orbit. 
        
        Specifically, we compare constellations featuring 2, 4 and 6 SpIRIT-like satellites in Polar orbit (plus HERMES - TP/SP), since these constellations ensure that either one, two or three Polar satellites are observing each of the co-pointing directions of the HERMES - TP/SP. For example, the localisation capabilities that we present for the $n_\text{polar} = 4$ constellation are a product of observation from 3 equatorial HERMES - TP/SP satellites plus 2 Polar SpIRIT-like satellites (see Figure \ref{fig:Constellation Pointing & Visibility}).

        \subsubsection{Long GRBs}
        \label{sec:Multi Polar long}
            
            Figure \ref{fig:Constellation CDF long} demonstrates that a constellation of four Polar satellites plus HERMES - TP/SP would improve the localisation capabilities of the constellation by a factor of $\sim 2$ over the case of a constellation with only one Polar satellite, and would be capable of localising $60\%$ of long GRBs to within $\sim 10$\,deg$^2$ of random uncertainty on the sky ($\sim 2\degree$ radius). A constellation of six Polar satellites would reflect a further improvement by an additional factor of $\sim 1.5$ (or equivalently a $4\times$ improvement over the case of a single Polar satellite), and could enable the localisation of $\sim 50\%$ of long GRBs to within a radius of $1\degree$ on the sky.
            
            Despite the substantial improvement over the case of a single Polar-orbiting satellite, GRB localisations with an error radius of $\sim 1\degree$ are not accurate enough to enable reliable afterglow follow-up observations from existing ground or space-based observatories, most of which have a FoV of the order $10 \times 10$ arcmin$^2$, e.g., GROND: $10' \times 10'$ \citep{Greiner_2008}, Swift UVOT: $17' \times 17'$ \citep{Troja_2020}, Gemini Multi-Object Spectrograph (GMOS): $5'.5 \times 5'.5$ \citep{Hook_2004}. Taking effective advantage of these GRB detections at high energy for infrared follow-up may, however, be possible in the future with $\gtrsim 1$ deg$^2$ FoV space-based telescopes with rapid repointing capabilities, such as the SkyHopper nano-satellite concept \citep{Thomas_2022}.
            
        \subsubsection{Short GRBs}
        \label{sec:Multi Polar short}
            
            As Figure \ref{fig:HERMES+SpIRIT short} demonstrates, launching a constellation of 4 or 6 Polar orbiting satellites into Polar orbit can improve short GRB localisations by a factor of $\sim 2-4$ over a constellation with a single Polar satellite, achieving a 90\% $1\sigma$ short GRB localisation range between $\sim 20-4100$ deg$^2$ (4 Polar) and $12-2850$ deg$^2$ (6 Polar) respectively. These improvements enable the statistical localisation of short GRBs with an accuracy close to that reported by Fermi GBM for a small fraction of the best-localised bursts, meaning that such an expanded nano-satellite constellation could be valuable in the near real-time localisation of EM counterparts to BNS coalescences, especially in cases where a short GRB is observed by both Fermi GBM and the nano-satellite constellation (which is likely, given Fermi GBM observes such a large fraction of the sky).
            
            At the time of writing, real time triggers and $1\sigma$ localisations for short GRBs can be obtained with accuracies $< 10$'s of square degrees (e.g., Fermi GBM \citep{Goldstein_2020}, the Neil Gehrels Swift Observatory \citep{Gehrels_2004, Troja_2020}, the Gravitational wave high-energy Electromagnetic Counterpart All-sky Monitor (GECAM; \citet{Zhao_2021}). To localise a substantial fraction of short GRBs to this accuracy with the current design of the HERMES instrument (and under the assumption that the satellites are uniformly distributed around equatorial and Polar orbits), at least 20 satellites divided evenly between equatorial and Polar orbit all observing the same field are required (i.e., combined detection from at least 10 HERMES instruments with large baselines in both the equatorial and Polar planes). On the other hand, the performance forecast for the localisation of long GRBs indicates that with an increase in the effective area of the detector, a constellation with a smaller number of elements would have a strong potential to provide state-of-the-art and cost effective all-sky monitoring and localisation of short GRBs/gravitational wave counterparts. A detailed investigation on the trade off between detector collecting area and number of constellation elements in the HERMES constellation is beyond the scope of this work (though the reader is directed to \citealt{Greiner_2022}, where a similar study for a hosted instrument on the Galileo constellation is presented). 
            
\section{Conclusion}
\label{sec:Conclusion}

    In this paper we characterise the GRB localisation capabilities of the SpIRIT + HERMES - TP/SP mini-constellation using temporal triangulation techniques. We design a simplified mission simulation based on an optimised mission scenario for the HERMES - TP/SP outlined in \citet{Sanna_2020} and specific assumptions regarding SpIRIT's orbit and pointing strategy in order to estimate the burst localisation capabilities of the combined constellation.
    
    We find that the SpIRIT + HERMES - TP/SP constellation is able to localise $60\%$ of long GRBs within a $1\sigma$ confidence region of $28$ deg$^2$ on the sky, and $90\%$ of long GRBs within $150$ deg$^2$. These confidence regions represent an improvement by a factor of $\sim 5$ over the HERMES - TP/SP alone, and are comparable to the statistical burst localisation capabilities of the Fermi GBM (excluding systematic uncertainties for all satellites). The increased performance of SpIRIT + HERMES - TP/SP compared to HERMES - TP/SP alone is due to SpIRIT's Polar orbit, which affords it a large baseline perpendicular to the equatorial orbit of the six HERMES - TP/SP satellites. We find that when SpIRIT is over the poles, the localisation capabilities of the combined constellation improve by a factor of $\sim 5$ compared to when SpIRIT is at the equator.
    
    The short GRB localisation capabilities of SpIRIT + HERMES - TP/SP reflect a similar improvement by a factor of $\sim 5$ over HERMES - TP/SP alone, with the combined constellation localising $60\%$ of short GRBs within $\sim 1850$ deg$^2$ and $90\%$ of short GRBs within $6000$ deg$^2$ (1$\sigma$ confidence regions). In fact, for some faint short GRBs the constraint on the source's position from statistical localisation techniques is so loose that constellation achieves more accurate GRB localisations when SpIRIT does not detect the burst, as this restricts the search region to the fraction of the sky that is obstructed by the Earth from SpIRIT's perspective at the time of the burst. The short GRB localisation capabilities of the constellation is greatly reduced compared to long GRBs due to the limited precision in the signal cross-correlation between satellites, which could be addressed by increasing the collecting area of the HERMES instrument. While this forecast performance is not competitive with state-of-the-art localisation of Fermi GMB, the combined SpIRIT + HERMES - TP/SP is still expected to provide a meaningful contribution to multi-messenger astrophysics. This will be both for events that might be missed by Fermi due to Earth occultation, as well as through the unique time resolution of the HERMES instrument, which would enable novel investigation into rapid variability of the high energy emission. 
    
    In order to further augment the capabilities of the HERMES - TP/SP, we investigated the localisation capabilities of an expanded nano-satellite constellation which includes additional satellites in Polar orbit. We find that a constellation covering two nearly orthogonal orbital planes comprising of 4 or 6 nano-satellites in each plane is capable of effectively localising $60\%$ of long GRBs within $\sim 5-10$ deg$^2$, and $60\%$ of short GRBs to within $\sim 550-850$\,deg$^2$, which may enable an expanded nano-satellite constellation to aid in the timely localisation of bright short GRBs/GW transients. Furthermore, a constellation with 6 elements in Polar orbit enables $50\%$ of long GRBs to be localised to within a statistical $68\%$ uncertainty radius of $\sim 1\degree$ on the sky, which may enable systematic afterglow follow-up observations from future instruments with FoV's $\geq 1$ deg$^2$. Furthermore, it will be possible to integrate the expanded HERMES nano-satellite constellation into the IPN in order to achieve even more accurate sky localisations for both long and short GRBs, though this improved localisation would only be available at later times since the IPN latency can be substantial.
    
    Therefore, after successful on-orbit demonstration of the concept that is expected to be achieved by the SpIRIT + HERMES - TP/SP satellites, augmenting further the elements in Polar orbit would be an ideal next step towards achieving an all-sky X-ray/gamma-ray monitor based on distributed aperture detectors to complement future gravitational wave detection campaigns.

\section{Acknowledgements}
MTh. acknowledges support from an Australian Government Research Training Program (RTP) Scholarship. This research is supported in part by the Australian Research Council Centre of Excellence for All Sky Astrophysics in 3 Dimensions (ASTRO 3D), through project number CE170100013. Parts of this research were supported by the Australian Research Council Centre of Excellence for Gravitational Wave Discovery (OzGrav), through project number CE170100004. GG acknowledges PRIN-MUR 2017 (grant 20179ZF5KS). JR acknowledges support by the European Union’s Horizon 2020 Programme under the AHEAD2020 project (grant agreement n. 871158).


\bibliography{references}

\end{document}